\documentclass[12pt,preprint]{aastex}

\usepackage{graphicx}
\usepackage{txfonts}
\usepackage{natbib}
\begin{document}
%


\title{Plasma heating in solar flares and their soft and hard X-ray emissions}


\author{R. Falewicz}
\affil{Astronomical Institute, University of Wroc{\l}aw, 51-622
Wroc{\l}aw, ul. Kopernika 11, Poland}
\email{falewicz@astro.uni.wroc.pl}





\begin{abstract}
In this paper, the energy budgets of two single-loop like flares observed in X-ray are analysed under the assumption that non-thermal electrons (NTEs) are the only
source of plasma heating during all phases of both events. The flares were observed
by \emph{RHESSI} and \emph{GOES} on February $\rm 20^{th}$, 2002 and June $\rm 2^{nd}$, 2002, respectively. Using a 1D hydrodynamic code for both flares the energy deposited in the chromosphere was derived applying \emph{RHESSI} observational data. The use of the Fokker-Planck formalism permits the calculation of distributions of the non-thermal electrons in flaring loops, thus spatial distributions of the X-ray non-thermal emissions and integral fluxes for the selected energy ranges which were compared with the observed ones. Additionally, a comparative analysis of the spatial distributions of the signals in the \emph{RHESSI} images was conducted for the footpoints and for the entire flare loops in selected energy ranges with these quantities fluxes obtained from the models. The best compatibility of the model and observations was obtained for the June $\rm 2^{nd}$, 2002 event in the 0.5-4 \AA\
 \emph{GOES} range and total fluxes in the 6-12 keV, 12-25 keV, 20-25 keV and 50-100 keV energy bands. Results of photometry of the individual flaring structures in a high energy range shows that the best compliance occurred for the June $\rm 2^{nd}$, 2002 flare, where the synthesized emissions were 30\% or more higher than the observed emissions. For the February $\rm 20^{th}$, 2002 flare, synthesized emission is about 4 times lower than the observed one. However, in the low energy range the best conformity was obtained for the February $\rm 20^{th}$, 2002 flare, where emission from the model is about 11\% lower than the observed one. The larger inconsistency occurs for the June $\rm 2^{nd}$, 2002 solar flare, where synthesized emission is about 12 times greater or even more than the observed emission.
Some part of these differences may be caused by inevitable flaws of the applied methodology, like by an assumption that the model of the flare is symmetric and there are no differences in the emissions originating from the feet of the flare’s loop and by relative simplicity of the applied numerical 1D code and procedures. No doubt a significant refinement of the applied numerical models and more sophisticated implementation of the various physical mechanisms involved are required to achieve a better agreement. Despite these problems, a collation of modelled results with observations shows that soft and hard X-ray emissions observed for analysed single-loop like events may be fully explained by electron beam-driven evaporation only.


\end{abstract}




\section{Introduction}
Despite the significant progress already achieved both in observational and theoretical
investigations as well as in numerical modelling of solar flares, the main physical processes
involved in energy conversion, transfer, deposition and losses in solar flares are only understand
fragmentarily.  However, broad insight is crucial for solving the most important questions, like a
balance of energy in the solar flares, a prediction of solar flare magnitudes and their emission
spectra, and emission of high energy particles. It is commonly accepted that during the impulsive
phase of solar flares as well as to some extent, before and after the impulsive phase, beams of
non-thermal electrons, accelerated somewhere in the solar corona stream along magnetic field lines
toward the chromosphere, where they heat by collision the dense matter near the feet of the loops.
The interactions of the non-thermal electrons with  dense matter also cause a strong and variable
in time emission of  hard X-rays (HXR) called electron-ion bremsstrahlung which was described by
\citet{Brown71} and \citet{Emslie78}. The chromospheric matter, heated up to coronal temperatures,
expands and fills the magnetic ropes (such a process is called chromospheric evaporation) and it
mainly emits soft X-rays (SXR) and also some thermal hard X-rays \citep{Anton84,Fis85b}. Taking
into account that the main physical processes occurring in solar flares are very complicated and
cross-related and very difficult in comprehensive analytical description, a numerical modelling of
the flares based on their observational properties allows us to extend our understanding of the
physical aspects of the flares. The first numerical models explored the hydrodynamic response of
the solar atmosphere to heating by electron beams
\citep{Nag84,MacNe84,Maris85,Fis85a,Fis85b,Fis85c,Maris89,Serio91}. \citet{Reale97} derived the
method of estimation of the sizes of spatially unresolved solar and stellar coronal flaring
structures based on the X-ray light curves and time-resolved temperature and emission measure values
measured during flare decay, applying hydrodynamic modelling. More sophisticated multi-thread,
time-dependent hydrodynamic simulations of solar flares was applied by \citet{Warr06} to
observations of the Masuda's flare of January $\rm13^{th}$, 1992.
Combined modelling of acceleration, transport, and hydrodynamic response in solar flares
was described in the paper by \citet{Liu09}. They tested three models of solar
flare heating, repeating the same simulation with two kinds of injection downward-beamed electrons using
different methods of particle transport and the new heating rates. As a main result of this
comparison, the model calculated using analytical approximation of particle transport gives about 10\%
difference compared to the model calculated with the Fokker-Planck formalism but provides an acceptable
approximation. In addition, it was found that using the model with the Fokker-Planck description of the particle transport
and if the injected electron spectrum is based on stochastic acceleration, the authors found higher
coronal temperatures and densities, larger upflow velocities, and faster increases of these
quantities than in the model with electron injection of power law.
\citet{Fal09a} using hydrodynamic modelling examined the sensitivity of the \emph{GOES} classification to the
non-thermal electron beam properties using \emph{Yohkoh} data, while a similar analysis was performed for the \emph{RHESSI} data by \citet{Reep13}.

In this paper we used all achievements of our previous
investigations of the energy transfer and deposition processes in simple, single-loop solar flares,
in particular our investigations of interactions of the non-thermal electrons with matter
\citep{Fal09a,Fal09b,Fal11,Siar09} and numerical modelling of the X-ray flare emission
\citep{Fal09b,Fal11}. Additionally, in our work we took into account the results presented
by \citet{Liu09}, which validated certain simplifications during the modelling of solar
flares, and were helpful in the interpretation of the results. We applied an essentially modified
by us hydrodynamic numerical code of the Naval Research Laboratory (NRL), see \citet{Mar82,Maris89}
and \citet{Fal09a} where a detailed description of the modified code is given. We modelled a heating process of the flaring plasma using Fisher's approximation \citep{Fis89} on the basis of energy flux
parameters evaluated using observed HXR spectra. In the present paper we apply the Fokker-Planck
formalism of the electron precipitation process through the plasma in the loops following method
already evaluated by \citet{McTier90}. We computed the distributions of the non-thermal electrons
in a flaring loop, spatial distributions of the X-ray non-thermal emissions and integral fluxes for
the selected energy ranges which were compared with the observed fluxes in relevant energy ranges.
Additionally, we also conducted a comparative analysis of the spatial distribution of the
signal in the \emph{RHESSI} images for the footpoints and entire flare loops in selected energy
ranges, and compared these quantities with the fluxes obtained from the models. A
conformity of fluxes was applied as a measure of the quality of the calculated flare models and
applicability of used procedures and simplifications.

In this paper we present the results of the numerical modelling of two solar single-loop like
flares. Section 2 describes the analysed events. Section 3 shows the details of the HXR spectra
fitting procedure, numerical modelling algorithm and methods of calculation of thermal and
non-thermal emissions. Section 4 presents the obtained results. The conclusions are discussed in
Section 5.

\section{Observational data}
For analysis we selected solar flares which look like single-loop structures in X-rays, best for
applied numerical model. The found flares were observed on February $\rm 20^{th}$, 2002, and June $\rm 2^{nd}$, 2002.
Both events were presented and analysed in several papers. The emissions of the flares were registered by the Reuven
Ramaty High Energy Solar Spectroscopic Imager (\emph{RHESSI}) satellite \citep{Lin02}. The
\emph{RHESSI} was designed to image solar flares in photons having energies from soft X-rays (3
keV) to gamma rays (17 MeV) with a high temporal and energy resolution and also to provide high
resolution spectroscopy in the same energy range. The \emph{RHESSI} has nine coaxial germanium
detectors with high signal sensitivity and allows a restoration of the images and spectra which are
very useful for investigation of the non-thermal emission of solar flares. In case of excessively
high photon fluxes, \emph{RHESSI} automatically limits the readings using attenuators, unfortunately
inserting also troublesome discontinuities of the recorded data \citep{Lin02,Hur02,Smith02}.
However, the both investigated flares were observed with the A1 attenuator permanently on.
\emph{GOES} X-ray data were also used. The \emph{GOES} (Geostationary Operational Environmental
Satellites) satellites since 1974 have continuously recorded integrated solar X-ray fluxes in two
energy bands 1-8 \AA\  and 0.5-4 \AA\ with 3 s temporal resolution \citep{Donnelly77}. To confirm
the likely single loop structure of the analysed flares, additional observations from the
Extreme-ultraviolet Imaging Telescope (\emph{EIT}) were also used. \emph{EIT} was installed on
board the Solar and Heliospheric Observatory \citep[\textit{SOHO};][]{Dela95} and provides
full-disk images taken in four bands: 171 \AA, 195 \AA, 284 \AA, and 304 \AA\ with 5 arcsec spatial
resolution.

\subsection{February $\rm 20^{th}$, 2002 solar flare}
This event was discussed already by \citet{Asch02} studying the height at which
the HXR emission originates, by \citet{Sui02} modelling the HXR emission, and
by \citet{Kruc02} who investigated the relationship between the brightnesses
of the footpoints. \citet{Ver05} studied physics of the Neupert effect and
\citet{Guo11} presented a new method to study in detail the temporal evolution
of thermal and non-thermal photon fluxes applied to this flare.

The C7.5 \emph{GOES}
class flare occurred in active region NOAA 9825 on February $\rm 20^{th}$, 2002 which was
located near the western limb of the solar disk. In this region on that day,
before the start of the analysed flare four events occurred, M4.2 (02:44),
M5.1 (05:52), C2.5 (07:41), M4.3 (09:46) and afterwards two C-class C4.5 (11:29)
and C2.9 (15:59). Seven flares were observed in the active region on that day. The
SXR (1-8 \AA) emission of the analysed flare recorded by \emph{GOES} started at 11:02 UT,
reached its maximum at 11:07 UT, and was observed up to 11:29 UT.
\begin{figure*}[t]
\begin{center}
\includegraphics[angle=0,scale=0.40]{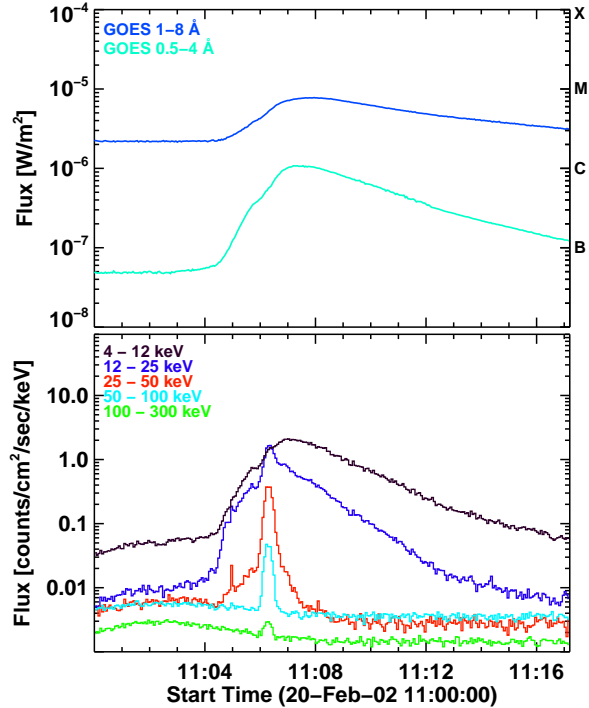}
\end{center}
\vspace{-1.0 cm}
\caption{\emph{GOES} X-ray 0.5-4 \AA\ and 1-8 \AA\ light curves (upper panel) and \emph{RHESSI} light curves of five energy bands between 4 and 300 keV (lower panel) taken during the
C7.5 \emph{GOES} class solar flare on February $\rm 20^{th}$, 2002.}
\end{figure*}
A harder emission recorded by \emph{GOES} (0.5-4 \AA) started to increase and peaked
thereabouts at the same time as the softer flux (1-8 \AA). \emph{GOES} light
curves of the analysed flare are presented in Figure 1 (upper panel). \emph{RHESSI} X-ray light curves of the flare taken in five energy bands are shown in Figure 1 (lower panel). The
impulsive phase of the flare registered by \emph{RHESSI} in X-rays above 25 keV started
at 11:04:28 UT, had its maximum at 11:06:08 UT and was observed even in the energy
range of 100-300 keV. The SXR emission registered by \emph{RHESSI} below 12 keV started
to rise simultaneously with \emph{GOES} 0.5-4 \AA\  emission and the time course of the
fluxes looks very similar.

\begin{figure*}[t]
\includegraphics[angle=0,scale=0.55]{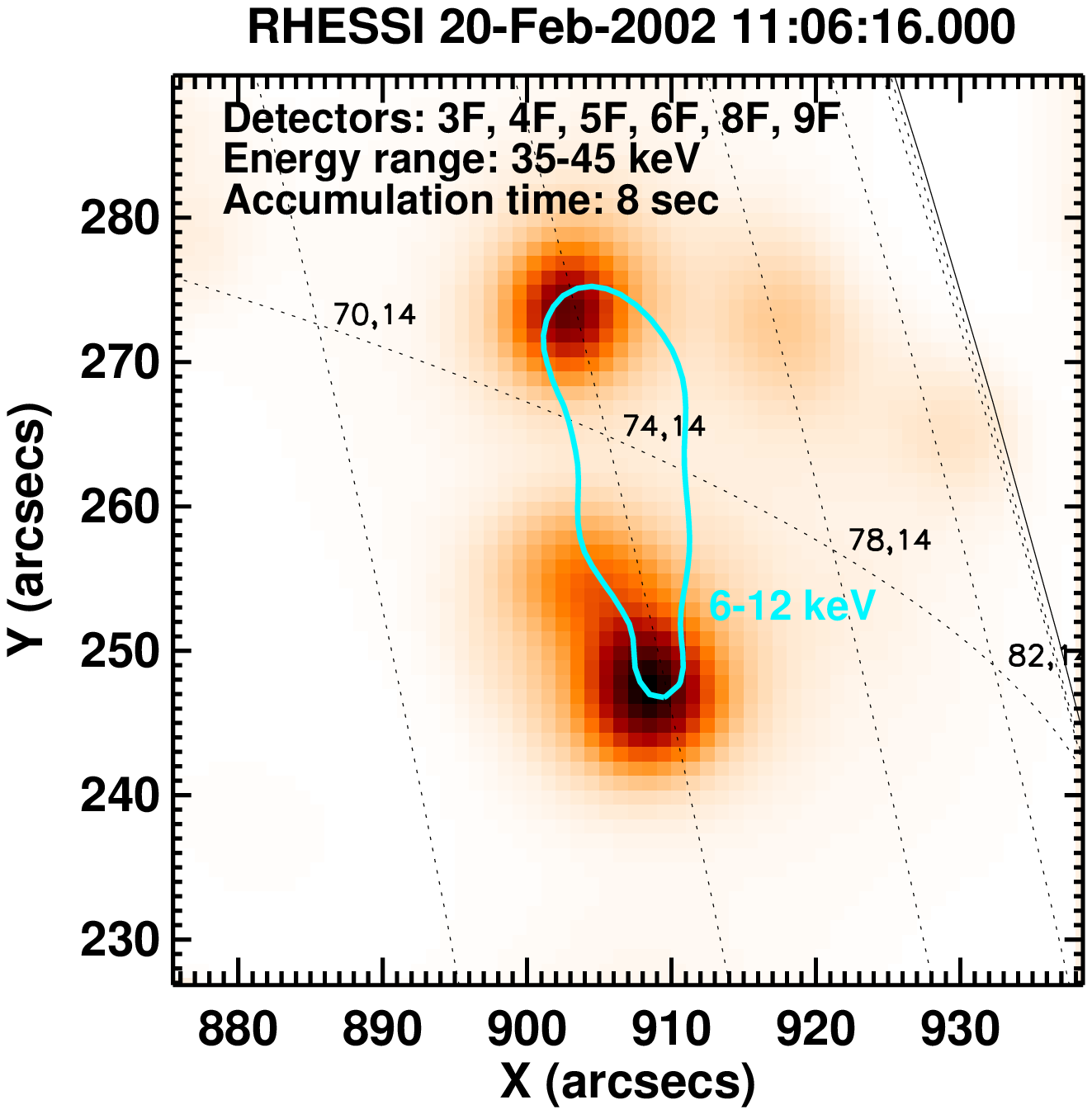}
\includegraphics[angle=0,scale=0.55]{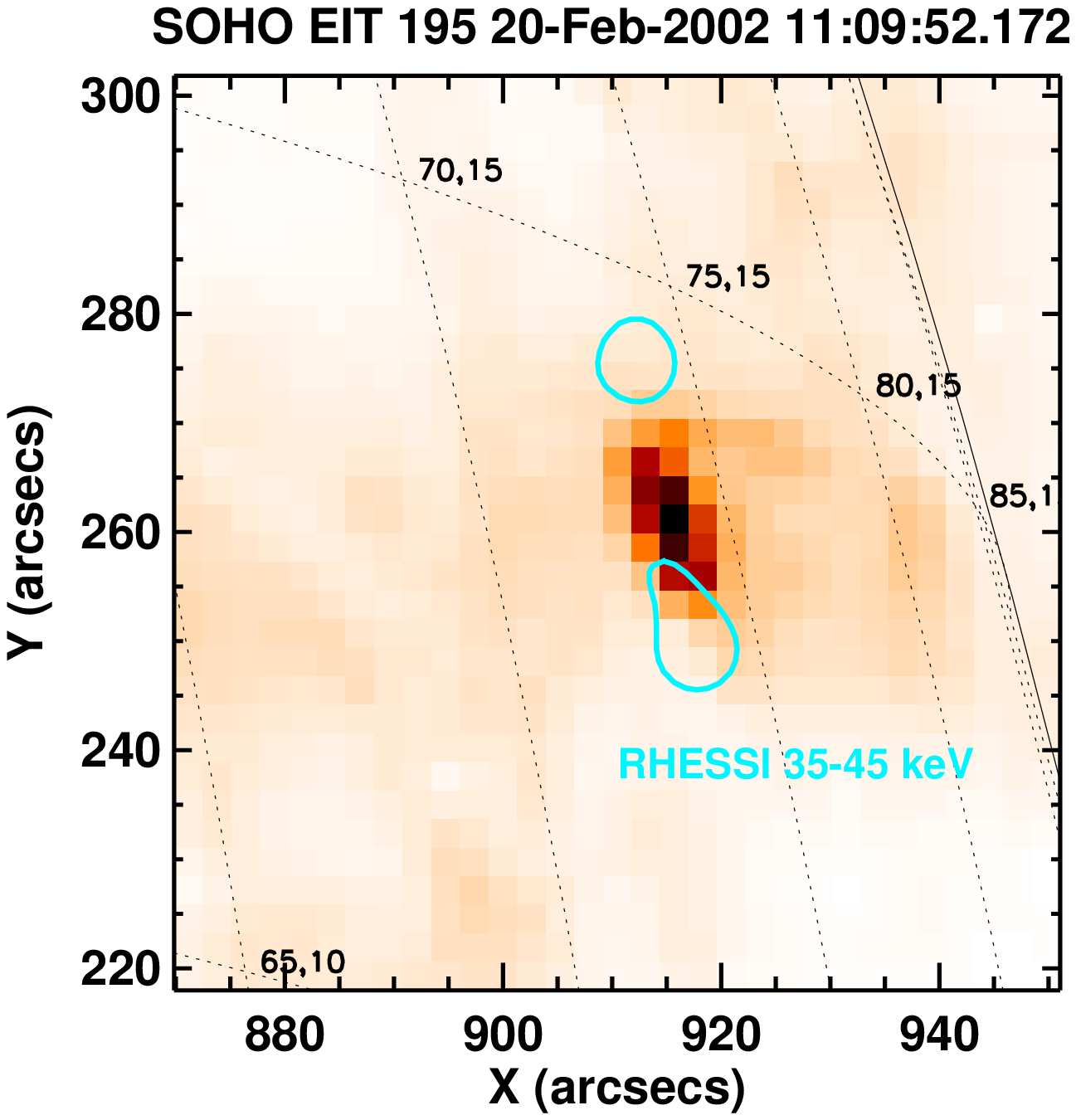}
\vspace{-0.5 cm}
\caption{Images of the C7.5 \emph{GOES} class solar flare on the February $\rm 20^{th}$, 2002.
Left panel: an image restored using the PIXON method in 35-45 keV energy band.
The signal was accumulated between 11:06:16 UT and 11:06:24 UT. Right panel:
\emph{SOHO/EIT} 195 \AA\ image taken at 11:09:52 UT (grey scale) overplotted with the
\emph{RHESSI} 35-45 keV image registered at 11:06:16 UT (contours).}
\end{figure*}

\begin{figure*}[t]
\includegraphics[angle=0,scale=0.34]{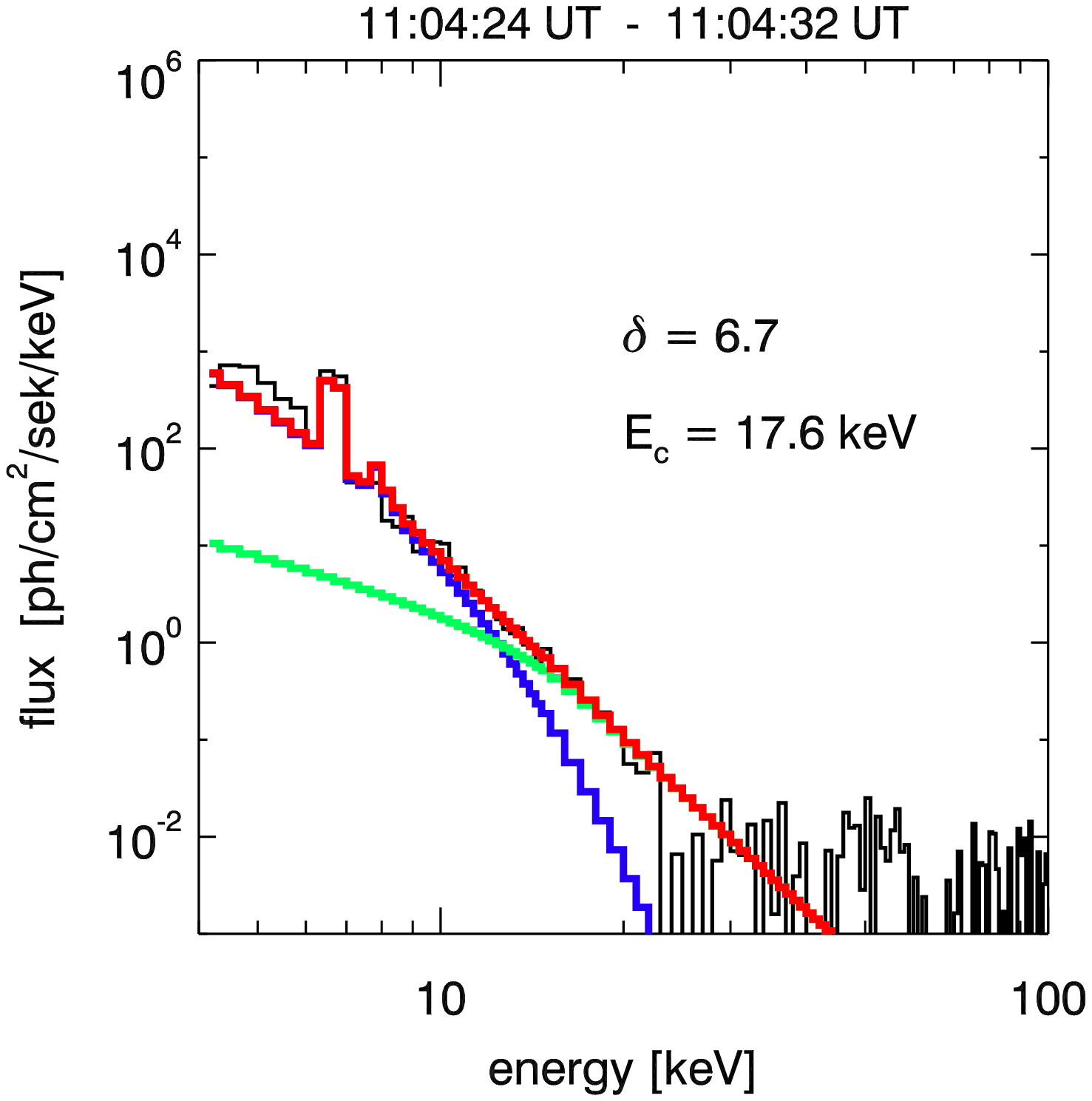}
\includegraphics[angle=0,scale=0.34]{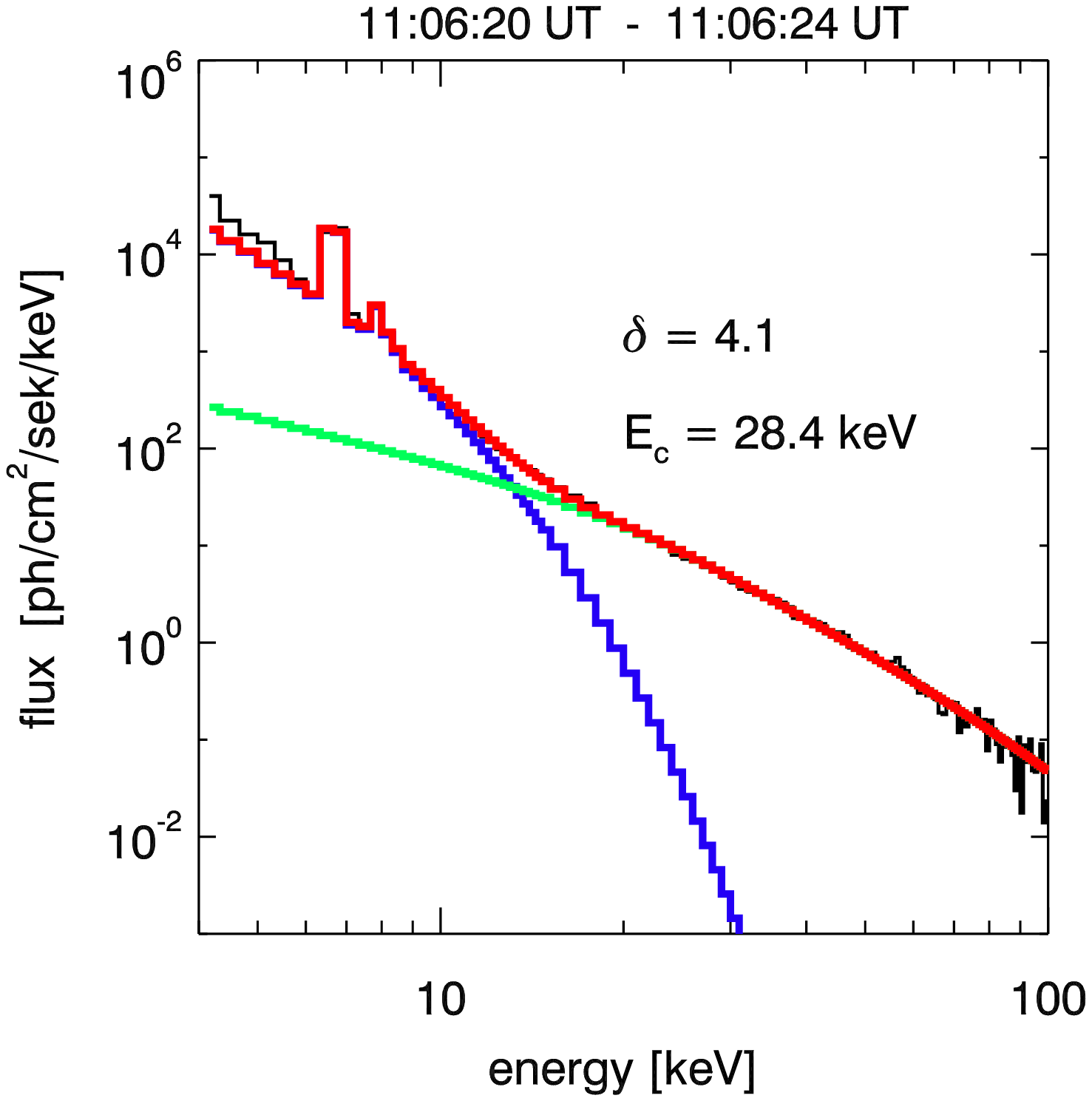}
\includegraphics[angle=0,scale=0.34]{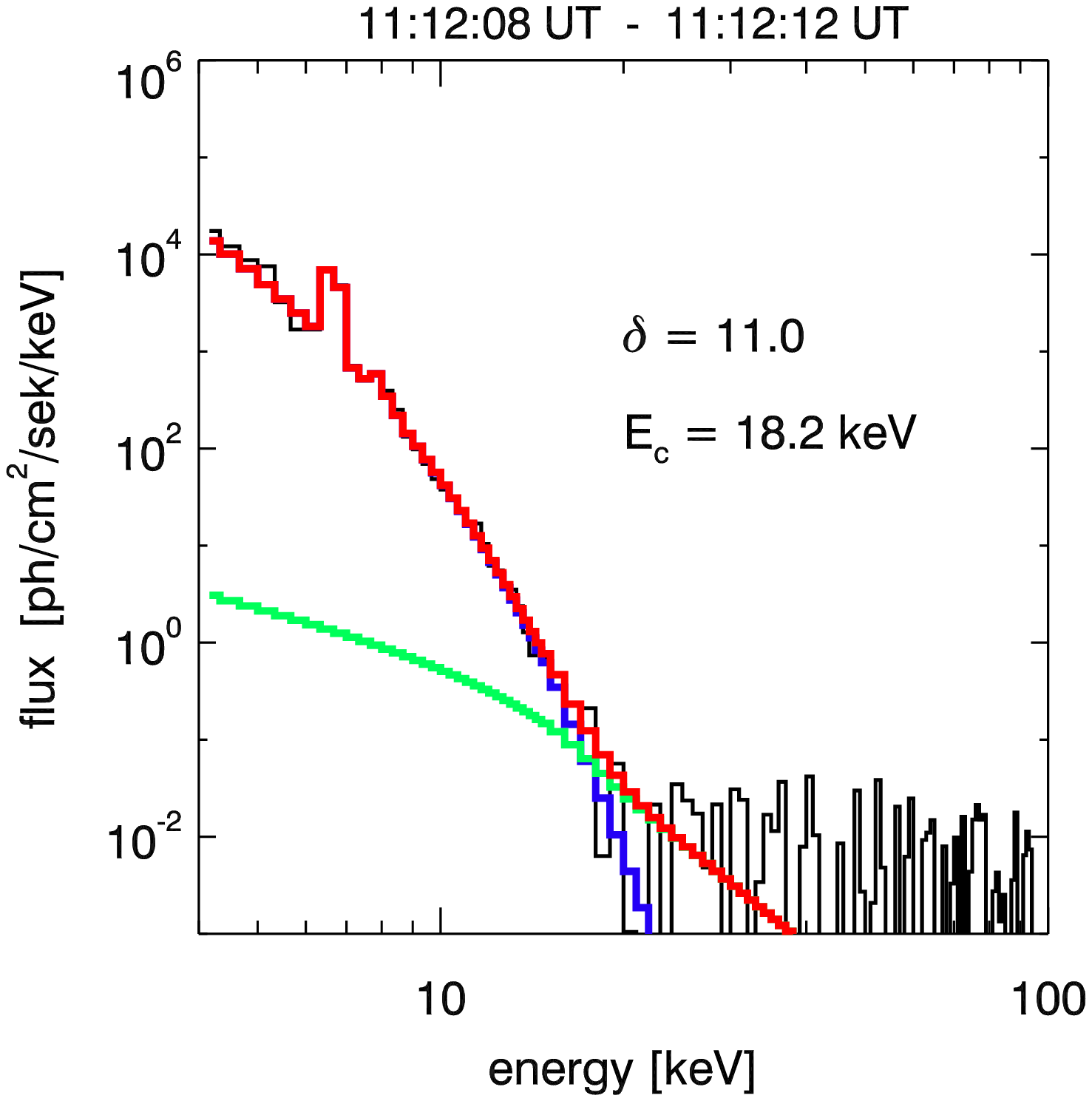}
\vspace{-0.5 cm}
\caption{\emph{RHESSI} spectra taken before (left panel), during (middle panel), and after the
impulsive phase (right panel) of the February $\rm 20^{th}$, 2002 flare. The spectra were fitted
with the single temperature thermal model (blue colour) and thick-target model (green). The summed
fits of the spectra are shown in red. The obtained values of the power-law index of the electron
energy distribution $\delta$, and low energy cut-off of the electron distribution $\rm E_c$ were
determined on the basis of the thick-target model and used as characteristics of the injected
electrons in the model.}
\end{figure*}

Images of the flare were reconstructed using \emph{RHESSI} data gathered with sub-collimators 3F,
4F, 5F, 6F, 8F and 9F integrated over 8 sec periods using PIXON imaging algorithm with a 1 arcsec
pixel size \citep{Met96,Hur02}. The images restored for the low energy band 6-12 keV (see Figure 2,
left panel) show a single source located between two footpoints visible in higher energy (35-45
keV). The image recorded by \emph{SOHO/EIT} 195 \AA\  at 11:06:15 UT (see Figure 2 right panel)
about four minutes later, indicates a structure very similar to single source recorded by
\emph{RHESSI} in 6-9 keV. Hard X-ray images were used to determine the main geometrical parameters
of the flaring loop using a method proposed by \citet{Asch99}. The cross-section of the loop $S=
(7.16\pm6.83)\times10^{16} \rm\, cm^2$ was estimated as an average area of both footpoints delimited
with isophotes of 30\% of the maximum flux in the 35-45 keV energy range. The cross-section of
flaring loops of both analysed events was assumed to be constant. Half-length of the loop
$L_0=(1.63\pm0.11)\times10^9$ cm was estimated from a distance between the centres of gravity of
the footpoints, assuming a semi-circular shape of the loop.  Samples of \emph{RHESSI} spectra taken
before, during, and after the impulsive phase of the February $\rm 20^{th}$, 2002 flare and fitted
with the single temperature thermal model and thick-target model are shown in Figure 3.

\subsection{June $\rm 2^{nd}$, 2002 solar flare}
The second analysed event was already described and analysed by \citet{Kruc02} who investigated the
relationship between the brightnesses of the footpoints. \citet{Ver05} analysed this flare in their
study of physics of the Neupert effect. In papers by \citet{Sui06,Sui08} the authors discussed the
morphology and evolution of the flare, particularly in the EUV and H$\alpha$ energy range. They
showed evidence for multiple-loop interactions as the cause of the flare. Also \citet{Mes09}
examined various scenarios of the flare and proposed a three-dimensional scheme in which the
filament eruption and flare were caused by interaction of the magnetic ropes.

\begin{figure*}[t]
\begin{center}
\includegraphics[angle=0,scale=0.40]{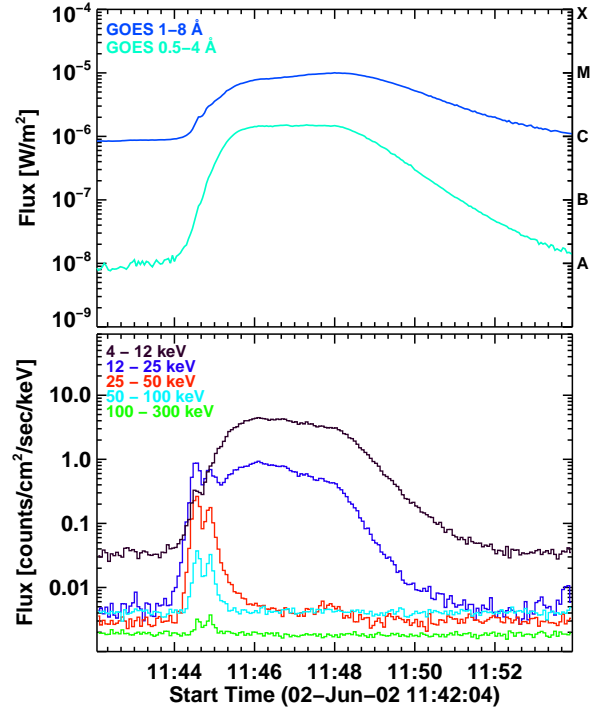}
\end{center}
\vspace{-1.0 cm}
\caption{\emph{GOES} X-ray 0.5-4 \AA\ and 1-8 \AA\ light curves (upper panel) and \emph{RHESSI} light curves of five energy bands between 4 and 300 keV (lower panel) taken during the C9.4 \emph{GOES} class solar flare on June $\rm 2^{nd}$, 2002.}
\end{figure*}

The investigated flare occurred in active region (AR) NOAA 9973 (S17E23) on June $\rm 2^{nd}$,
2002. It was classified as a C9.4 \emph{GOES} class flare. A magnetic class of the active region
was beta-gamma and it produced three C class flares on that day, including the analysed event. The
\emph{GOES} 1-8 \AA\ flux started to increase slowly at 11:41 UT, reached a maximum at 11:47 UT,
and was observed up to 11:55 UT. The time course of its SXR emission is not typical, having a flat
maximum particularly visible in harder energy band recorded by \emph{GOES} (0.5-4 \AA). \emph{GOES}
X-ray light curves of the flare are shown in Figure 4 (upper panel).  \emph{RHESSI} X-ray light
curves of the flare taken in five energy bands are shown in Figure 4 (lower panel). The impulsive
phase started in X-rays above 25 keV at 11:44:14 UT, had double maximum at 11:44:34 and 11:44:54
UT, and was observed up to 11:46:02 UT. The hard X-ray emission of the impulsive phase was observed
up to 300 keV. The X-ray emission below 25 keV started to rise simultaneously with the \emph{GOES}
emission. Small increases of emissions causing flattening of the SXR maxima occurred at the same
time in the \emph{GOES} emission and \emph{RHESSI} channels 4-12 keV and 12-25 keV at 11:47:42 UT.

\begin{figure*}[t]
\includegraphics[angle=0,scale=0.55]{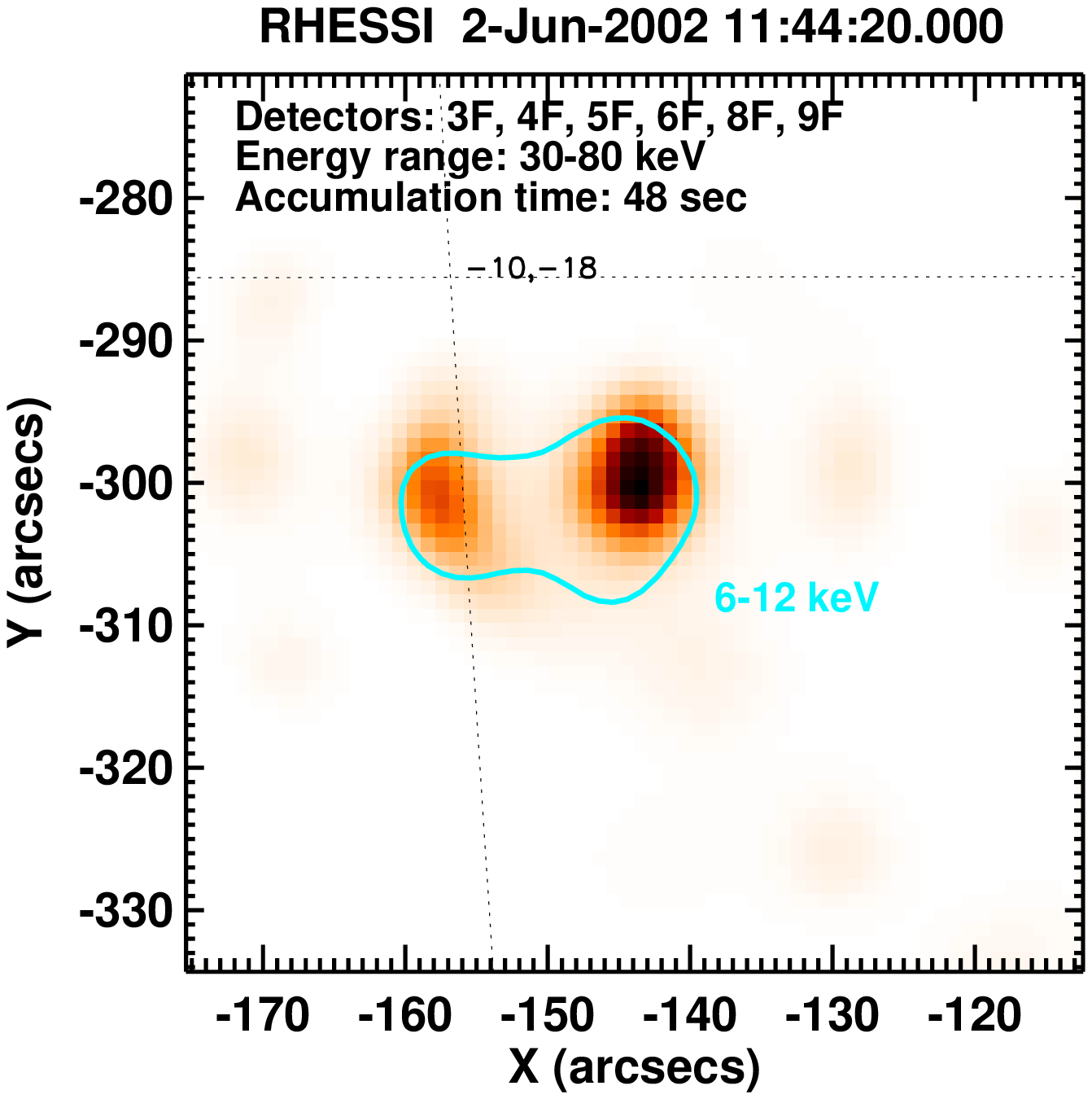}
\includegraphics[angle=0,scale=0.55]{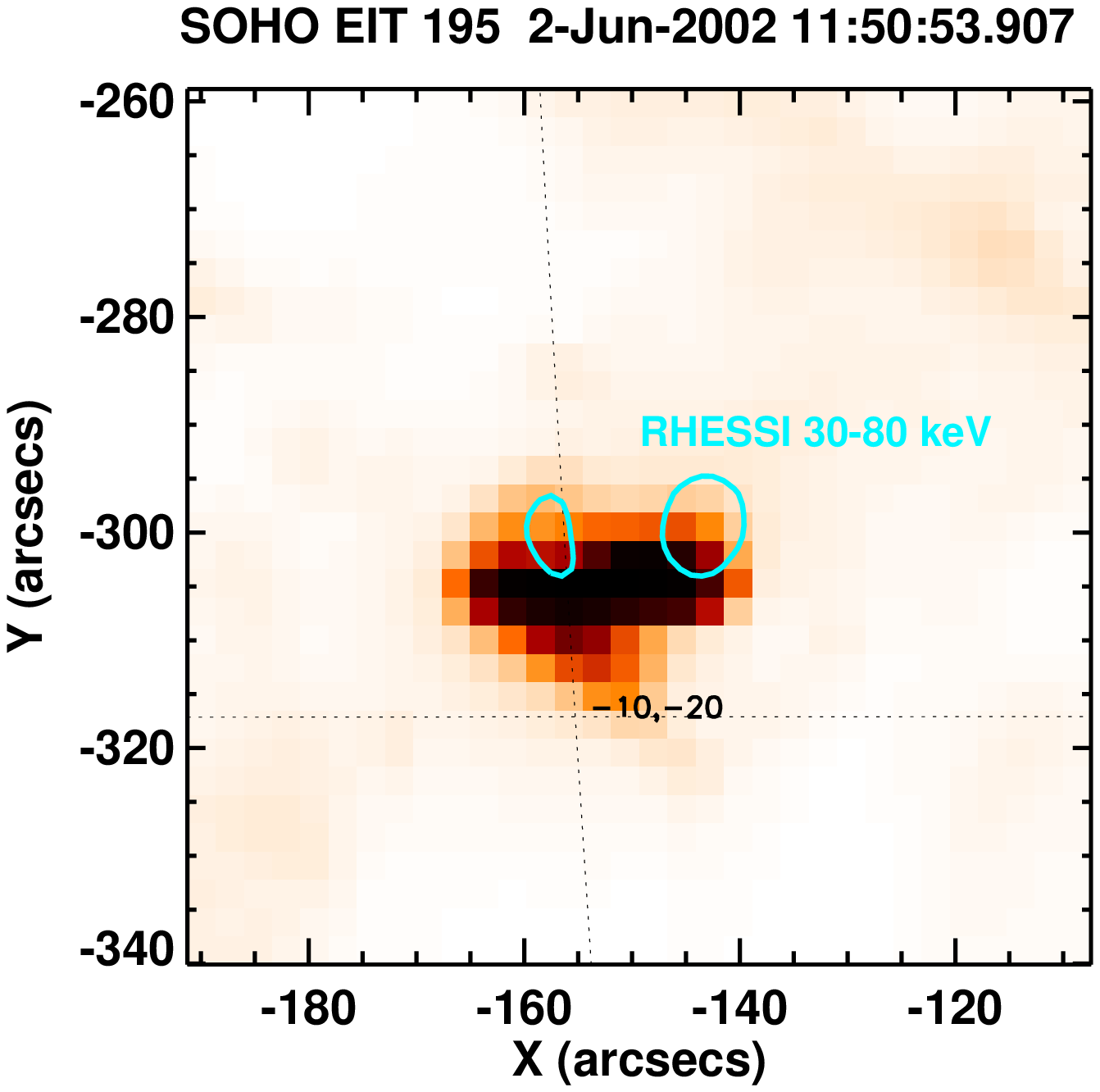}
\vspace{-0.5 cm}
\caption{Images of the C9.4 \emph{GOES} class solar flare on the June $\rm 2^{nd}$, 2002. Left
panel: an image restored using the PIXON method in the 30-80 keV energy band, the
signal was accumulated between 11:44:20 UT and 11:45:08 UT, at the maximum of
the impulsive phase (grey scale) overplotted with the \emph{RHESSI} 6-9 keV image
(contour). Right panel: \emph{SOHO/EIT} 195 \AA\ image taken at 11:50:54 UT, after the
impulsive phase of the flare, overplotted with the \emph{RHESSI} 30-80 keV PIXON image
(contour) registered at 11:44:20 UT.}
\end{figure*}

The start of the impulsive phase occurred simultaneously with an increase of the soft X-rays
emissions and no preheating prior to the impulsive phase was observed.

\begin{figure*}[t]
\includegraphics[angle=0,scale=0.34]{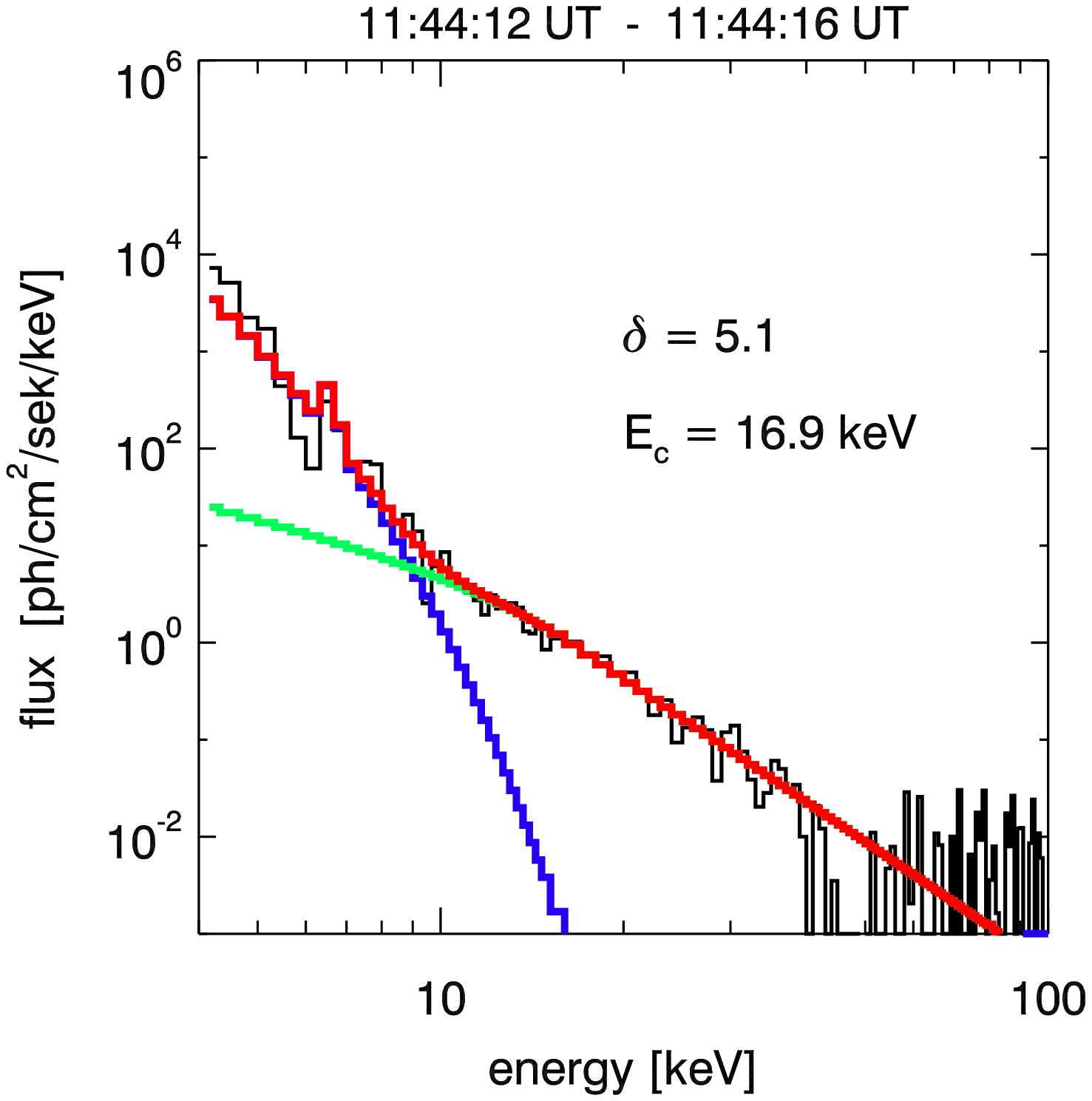}
\includegraphics[angle=0,scale=0.34]{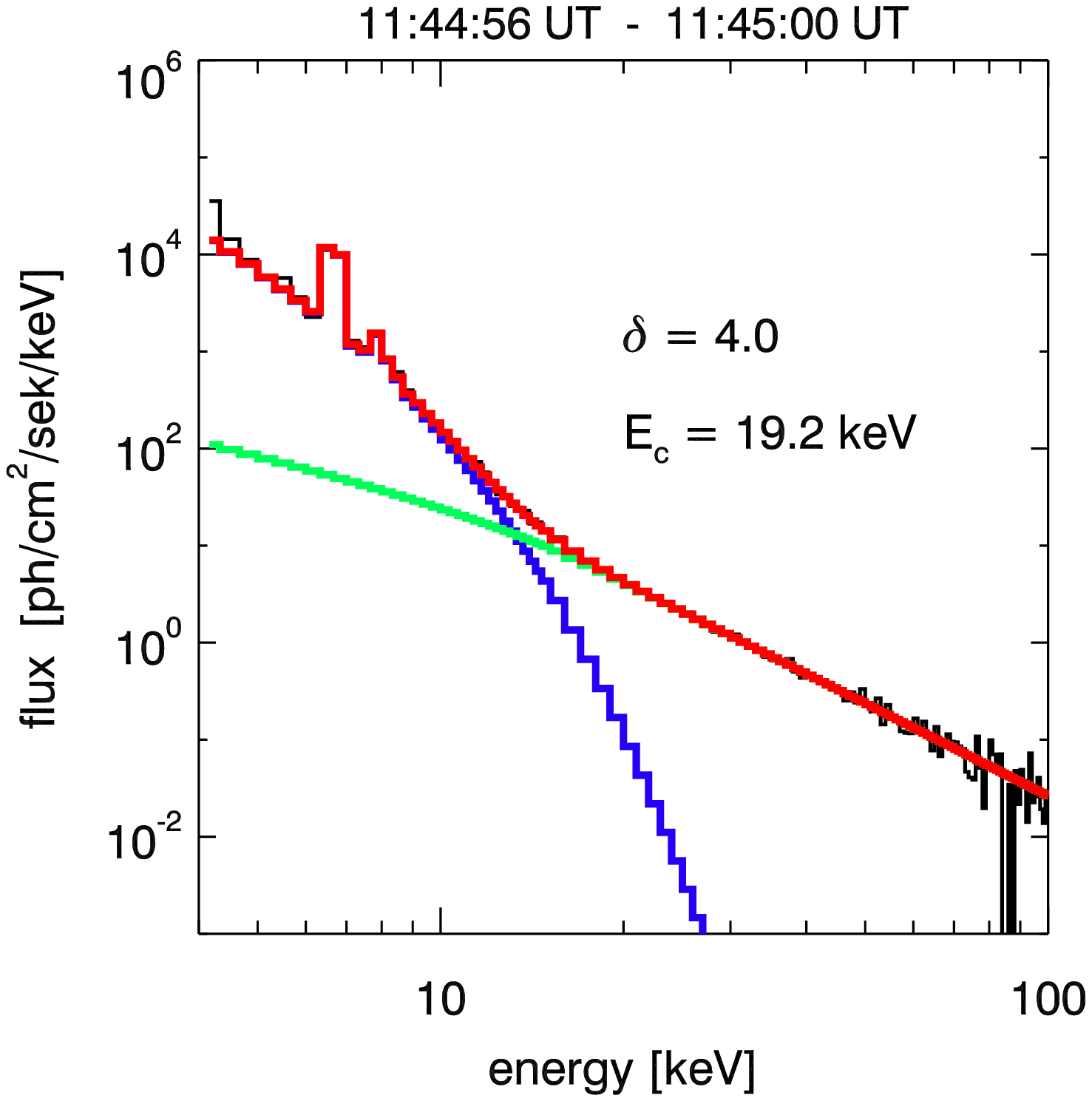}
\includegraphics[angle=0,scale=0.34]{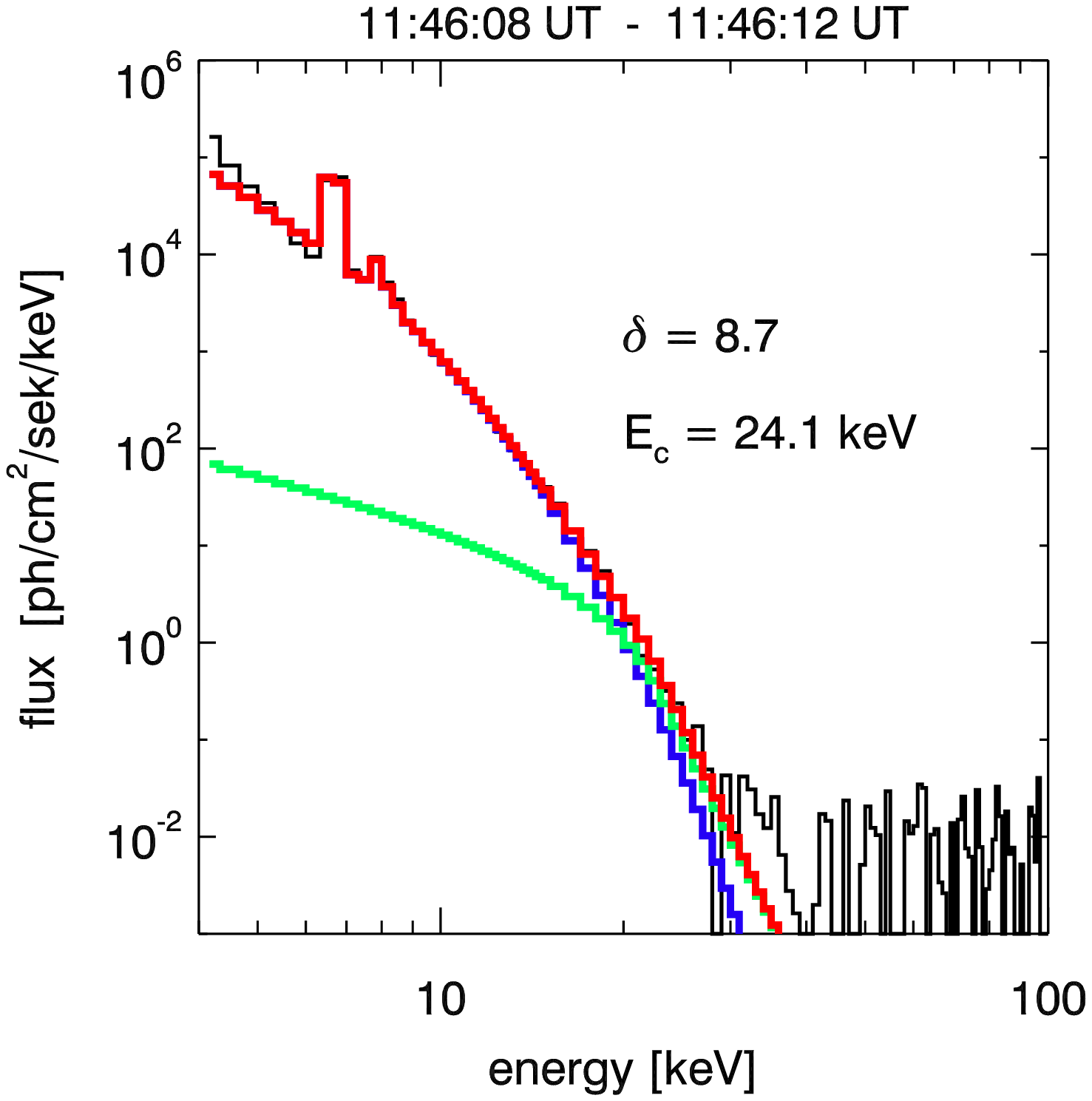}
\vspace{-0.5 cm}
\caption{\emph{RHESSI} spectra taken before (left panel), during (middle panel), and
after the impulsive phase (right panel) of the flare on the June $\rm 2^{nd}$, 2002. The
spectra were fitted with the single temperature thermal model (blue colour) and
thick-target model (green). The total fitted spectra are shown in red. The obtained values
$\delta$, and  $\rm E_c$ were determined on the basis of the thick-target model and used as
characteristics of the injected electrons in the model.}
\end{figure*}

The images of the flare were obtained using \emph{RHESSI} data collected with sub-collimators 3F,
4F, 5F, 6F, 8F and 9F integrated over 48 s periods and the PIXON imaging algorithm with 1 arcsec pixel
size. The image restored in the 30-80 keV energy range revealed two distinct footpoints, where the
western footpoint is brighter than the eastern (Figure 5 left panel).  In the 6-12 keV energy range
the entire flare loop was seen. The image recorded by \emph{SOHO/EIT} 195 \AA\ at 11:50:54 UT,
after the impulsive phase of the flare and overplotted with the \emph{RHESSI} 30-80 keV PIXON image
(contour) registered at 11:44:20 UT is shown in Figure 5 right panel. The main flare loop visible
in the EUV range spatially coincides well with footpoints visible in the \emph{RHESSI} 30-80 keV
range. A method of determination of the geometrical parameters was the same as for the first flare.
The cross section of the loop $S=(6.04\pm5.60)\times10^{16}\rm cm^{2}$ was estimated as an average
area of the feet delimited with isophote of 30\% of the maximum flux in the 30-80 keV energy range.
Half-length of the loop was estimated as $L_0=(8.37\pm1.13)\times 10^8$ cm. Samples of
\emph{RHESSI} spectra taken before, during and after the impulsive phase are shown in Figure 6.

\section{Calculations}
The procedure of the numerical modelling used in this paper is very similar to the one described in
previous publications \citep{Siar09,Fal11}. The main assumption of the method is that interactions
of the non-thermal electrons (NTEs) with plasma are the only source of plasma heating during all
phases of the analysed events. The same hydrodynamic model of the analysed flares was used, where
the non-thermal electron beams derived from \emph{RHESSI} spectra as the carriers of heating energy
deposited, via Coulomb collisions. The energy deposition rates were calculated using an
approximation given by \citet{Fis89}. As in the previous paper \citep{Fal11} the low energy cut-off
of the electron distribution $\rm E_c$ was automatically optimised by comparison at each time step
of the observed SXR \emph{GOES} flux in the 1-8 \AA\ range and the calculated one. In this paper in
addition, for each time step the steady-state electron spatial and spectral distributions in the
loop were calculated using a Fokker-Planck code \citep{McTier90}. Known distributions of the
non-thermal electrons and physical parameters of the plasma along the flaring loop allowed for the
calculation of spatial distributions of the X-ray thermal and non-thermal emissions, to derive
integral fluxes in the selected energy ranges and to compare them with the fluxes observed in the
same energy ranges (see a detailed description below).
Furthermore, a comparative analysis of the spatial distributions of the
signals in the RHESSI images and models of both flares was made.
We compared signals emitted from selected structures and within selected
energy ranges. The conformity level was applied as a
measure of the model quality and validity of the applied procedures and simplifications.

\subsection{Modelling of the flares}
The \emph{RHESSI} data were analysed using the \emph{RHESSI} OSPEX
package of the SolarSoftWare (SSW). The X-ray spectra of both flares were
measured with a 4 seconds time resolution in 158 energy bands ranging from
6 to 300 keV. The analysed spectra were corrected for pulse pileup, decimation,
and albedo effect (Compton back-scattering on the photosphere,
\citet{Bai78}) using the \emph{RHESSI} standard analysis tools.

The accuracy of background subtraction is a key point in spectra analysis. In the case of the
\emph{RHESSI} spectra background is usually modelled as a linear (or higher-order polynomial)
interpolation between averaged fluxes before and after the flare. However, for many flares and
periods between them gradual rises and falls of this background were observed. Such a slow increase and
then decrease of the flux were recorded for the February $\rm20^{th}$, 2002 flare before 11:04 UT.
Similar background variations were observed in all energies, also in the highest energy channel
where no flare emission was observed. Averaged time profiles of the high energy backgrounds were
used for the modelling of the background variation in the low energy channel during this flare.
This method allows for a subtraction of the backgrounds with sufficient accuracy in different energy
channels. For the June $\rm 2^{nd}$, 2002 event the background varied marginally with time during this
flare and was subtracted from data using a linear fit of the background fluxes recorded before and
after the flare.

The \emph{RHESSI} A1 attenuator was permanently used
during both analysed events. Thus, counts registered in the 3-6 keV energy range are almost exclusively
generated by higher energy photons above 11 keV, which is caused by the K-escape effect
\citep{Smith02}. Therefore, only counts above 6 keV were analysed.  For the February $\rm20^{th}$, 2002
event above 90 keV and for the June $\rm 2^{nd}$, 2002 above 100 keV the X-ray background flux
dominates over flare emission, so fluxes above these energy levels were also not included for
spectral fitting.

Particularly at the beginning and end of the flare when recorded fluxes
are low, the count rates in some energy bins can be negative as a result of
background subtraction and low signal to noise ratios. In order to keep the
count rates positive (particularly in the 6-20 keV energy range) an increased
accumulation time was applied, causing a worse time
resolutions for the beginnings and ends of both analysed events than maxima phases.
The spectra were fitted assuming single
temperature thermal plus thick-target models. The thermal model was
defined by a single temperature and an emission measure of the optically thin
thermal plasma, based on the X-ray continuum and line emissions calculated using the
CHIANTI atomic code \citep{Dere97,Landi06}. The thick-target
model was defined by the total integrated non-thermal electrons flux $\rm F_{nth}$, the
power-law index of the electron energy distribution $\delta$, and low energy cut-off of
the electron distribution $\rm E_c$. These parameters were then used as characteristics of the injected electrons in the model.
Figures 3 and 6 show examples of spectra and fits
the thermal and non-thermal component for analysed events and determined on the
basis of parameters of the thick-target model. The \emph{RHESSI} spectra were fitted
using a forward and backward automatic fitting procedure, starting from a time
when the non-thermal component was strong enough and clearly visible. The
obtained values of the fitted parameters were controlled and corrected
if necessary. Generally, all fitted parameters evolved in a quasi-continuous manner.

Geometrical and physical models of the pre-flare loops were built using geometrical parameters - semi-length $\rm L_0$ and cross-section S and also thermodynamic parameters like
initial pressure at the base of transition region $\rm P_0$, temperature, emission
measure, mean electron density and \emph{GOES} flux in the energy range 1-8 \AA\  were estimated from \emph{RHESSI} and \emph{GOES} data.  In order to
obtain for initial pre-flare models the best conformity between theoretical and
observed \emph{GOES} light curves, the semi-lengths and cross-sections of the loops
were refined in a range of $\pm$1 pixel. Table 1 presents the values of S, $\rm P_0$ and
$\rm L_0$ used in the models of the analysed flares.

\begin{table*}[t]
\caption{Cross-sections, semi-lengths and base pressure of the analysed flares.} 
\label{table:1} 
\centering 
\begin{tabular}{c c c c c c c c} 
\hline\hline 
Event     & \multicolumn{2}{c}{Time of}& \emph{GOES}&Active & S              & $L_0$        & $P_0$       \\ %
date      &start     & maximum         & class      &region &                &              &             \\
          & [UT]     &[UT]             &            & AR    &[$10^{16} cm^2$]& [$10^{8}cm$] & [$dyn/cm^2$]\\
 \hline
20-Feb-02 & 11:02    & 11:07           & C7.5       &9825   &7.16            & 17.2          & 63.0        \\
02-Jun-02 & 11:41    & 11:47           & C9.4       &9973   &5.81            & 8.37          & 72.0        \\

\hline 
\multicolumn{8}{l}{{\footnotesize $\rm S$ and $\rm L_0$ - cross-section and semi-length of the flaring loop; $\rm P_0$ - pressure at base of transition region}} \\
\vspace{-0.6cm}
\end{tabular}
\end{table*}

For each time step the momentary heating rate of plasma along the loop was estimated using Fisher's
heating function. In this function thick-target parameters $\rm F_{nth}$, $\delta$, and $\rm E_c$
of the NTEs beams obtained from fitting consecutive \emph{RHESSI} spectra were used. The time step
of the NTEs beam was defined as the accumulation time needed to obtain the spectrum from
\emph{RHESSI} data with a sufficient S/N ratio. For this reason, during each NTEs beam's time step
a heating of the loop was calculated with fixed NTE beam parameters, while the thermodynamic
parameters of the flaring plasma varied, being calculated with a much shorter time step of the HD
model. The best beam time resolution was obtained when the flux was the highest around the maximum
of event (being equal to 4 seconds), and the worst for the beginning and the end (a dozen or
several dozen seconds). At the end of each NTE's beam time step \emph{GOES} fluxes were calculated
and compared with the observed \emph{GOES} fluxes. If a conformity of the fluxes was achieved, the
procedure was applied for the next time step. Otherwise, the $\rm E_c$ value was carefully adjusted
in order to achieve conformity of the observed and modelled \emph{GOES} fluxes in the 1-8 \AA\
energy band.  A variation in the $\rm E_c$ value of just a few keV can add or remove a substantial
amount of energy to or from the flare loop because of the power-law nature of the energy
distribution. Thus $\rm E_c$ must be selected with high precision. Usually the value of $\rm E_c$
obtained from fitting \emph{RHESSI} spectra is fraught with a large error. However the applied
method limits the range of $\rm E_c$ using an independent energetic condition, like the observed
1-8 \AA\  \emph{GOES} flux.

\subsection{Calculation of thermal emission}
The thermal emission fluxes were calculated using the CHIANTI atomic code \citep{Dere97,Landi06}
for the temperature range 1-50 MK, ionization equilibrium by \citet{Mazz98} and solar coronal
extended abundances. Thermal emission was calculated for both the lines and continuum. The
procedure for calculating the thermal emission fluxes consisted of several stages. In the first,
thermal emission was calculated for each cell of the model and in four energy ranges: 6-12 keV,
12-25 keV, 25-50 keV and 50-100 keV at the end of each NTE's beam time step. Next, the total flux
was calculated for the entire loop in the same energy ranges. As a result the light curves in
different energy ranges and spatial distribution of the thermal emission were obtained.

\subsection{Calculation of non-thermal electron distribution}
In this work, the Focker-Planck formalism was applied using a code originally developed by
\citet{Leach81}, based upon a program developed by Walt, MacDonald, and Francis \citep[described
in][]{Caro68}, extended to ultrarelativistic energies by \citet{McTier90} and prepared for public
distribution by Holman (\emph{http://hesperia.gsfc.nasa.gov/hessi/\\ flarecode/efluxprog.zip}). This code
computes the steady-state electron distribution function in the flare's magnetic loop as a function
of a position along a magnetic field line, electron pitch angle, and electron energy. Input data
consist of injected flux, distribution of non-thermal electrons at the top of the loop, a model of
a distribution of plasma in the flare's loop and a model of the magnetic field. The solution
includes effects due to Coulomb collisions with the ambient plasma, effects due to synchrotron
emission by the electrons, and effects of bremsstrahlung emission. It should be noted that
most of the non-thermal electrons lose their energy in Coulomb collisions while a small fraction ($\thicksim10^{-5}$)
of the electrons' energy is converted into HXR by the bremsstrahlung process.
In addition, a synchrotron effect is insignificant in solar flares. An
assumption of a steady-state imposes that the injected flux of NTEs is constant in time for all
time scales smaller than the transport time scales. In other words, the time of flight of the NTEs
from top of the loop to bottoms (feet) must be shorter than the interval, when parameters of the
injected NTEs beam at the top of the loop varied. The code has been modified to work with
hydrodynamic code in order to retrieve physical parameters of each cell along a computational loop.
As a result the procedure calculates the electron flux distribution $\rm F(E,s,\mu)$ as a function
of energy E, at all points in the loop (distance from the top of loop s) and for all angles (pitch
angles $\mu$=cosine $\theta$) in electrons/$\rm cm^2/sr/sec/keV$. For both analysed
flares, constant cross-section of loops was assumed \citep[for what seems to be a good
approximation, see][]{Peter12} which implies no magnetic mirroring because magnetic field strength
was constant and equal to 200 Gauss. The value has been selected on the basis of estimates of the
coronal magnetic field presented in \citet{Schmeltz94} and \citet{Asch05}.

\subsection{Calculation of non-thermal emission}
Known distribution of NTEs along the loop allows for the calculation of the thin-target non-thermal
bremsstrahlung radiation intensity $I(\varepsilon,s)$ at the end of each time step. The intensity
is a function of photon energy $\varepsilon$ and distance from the top of loop s, and by using the
formula of \citet{Tand88} one has:

\begin{equation}
I(\varepsilon ,s)=S\int_{\varepsilon}^{\infty} n_p(s) \sigma_B F(E,s)dE
\label{eq1}
\end{equation}

where S is a cross-section area of a loop, $\rm n_p$ is the ambient proton density as a function of
the distance along the injected electron's path for full ionization. $\rm \sigma_{B}$ is the
angle-averaged differential bremsstrahlung cross-section given by \citet{Koch59}. Electron flux
distribution $F(E,s)=\int_{-1}^{1} F(E,s,\mu)\, d \mu$ was integrated over the pitch angle from 1
to -1. The $I(\varepsilon,s)$ is calculated in $\rm ph/cm^{2}/sec/keV$. As a result the light
curves and spatial distribution in different energy ranges of the non-thermal HXR emission is
obtained.

\subsection{Observed and synthesized light curves}
Observed hard X-ray light curves in ph/$\rm cm^2$/sec/keV were restored on several assumptions: the
\emph{RHESSI} attenuator state during the analysed flares was permanently A1, the resulting fluxes
were obtained by dividing the instrument response matrix with diagonal coefficients only, thus
other factors were not taken into account e.g. there are no corrections for effects such as
K-escape and Compton scattering (described by the off-diagonal elements of the detector response
matrix). Both analysed flares were in the upper range of \emph{GOES} class C (C7.5 and C9.4), which
does not emit high hard X-ray fluxes. The correction of decimation and pileup was also included.
Observed light curves were prepared in four energy bands: 6-12 keV, 12-25 keV, 25-50 keV and 50-100
keV. To make a comparison of modelled and observed fluxes possible, it was necessary to add to each
of them the HXR thermal and non-thermal synthesized component and then add a background set of
observational light curves. In this way it was possible to compare the synthesized and observed
fluxes in the same energy range selected. Figure 7 shows a schematic flow diagram used for the
modelling of solar flares and their analysis. All of the components in the diagram are described in
this section.

\begin{figure*}[t]
\begin{center}
\includegraphics[angle=0,scale=0.50]{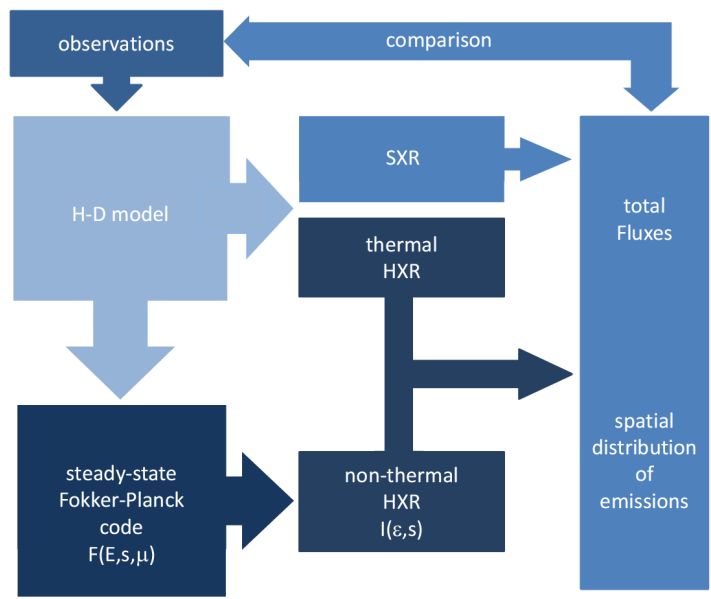}
\end{center}
\vspace{-0.5 cm}
\caption{Flow diagram showing the procedures used during modelling and analysis of flares.}
\end{figure*}

\section{Results}
The main results of modelling the analysed flares are presented in Figures 8 and 9 for June $\rm
2^{nd}$, 2002 and Figures 10 and 11 for Feb $\rm 20^{th}$, 2002. The synthesized \emph{GOES} 1-8
\AA\  light curves of the flaring loops follow closely those observed for the analysed flares,
which is a direct result from our assumptions. The real test for the accuracy of the model is to
compare the observed and calculated fluxes in the 0.5-4 \AA\  band. The quality of the fit between
an observed and synthesized light curve in the 0.5-4 \AA\  band would be a good measure of quality
of the calculated flare model. In other words, with smaller differences between light curves in the
0.5-4 \AA\  band, one gets a better model/simulation, larger differences, and a worse
model/simulation.  For the event of June $\rm 2^{nd}$, 2002 one can trace this effect in Figure 8
(upper left panel). The calculated light curves for the analysed event did not differ too much from
the observed ones. It seems that our model simulates the main physical processes in the right way
despite the large simplifications which are used. At the beginning of the flare the synthesized
light curve (red colour) is under observed (blue dashed line), after 11:44:20 UT have a similar
pattern to the maximum at 11:45:40 UT. After this time, during the flat maximum the synthesized
curve is located again under the observed one to 11:48:20 UT. Again, after this time the curves are
in a similar course to the end of the simulation. Additionally, one can see that during the decay
phase (after 11:48:20 UT) some heating is necessary. The green dotted curve in Figure 8 (upper left
panel) represents the evolution of the plasma in the flaring loop without any heating and has a
different course to the observed one. The observation corresponds well to the evolution of plasma
flares in the loop which is heated in this time.  A difference in the course between synthesized
and observed light curves also reflects on the calculated temperature and emission measure values.
We can follow the behaviour temperature with time in Figure 8 (upper right panel). Obviously, where
there are the greatest differences in the flux 0.5-4 \AA\  energy band, there will also be the
largest differences between the temperatures determined from observations and synthesized curves.
From almost the beginning of the flare, the temperature is lower than that determined at
the same time from observations but both values agree within the range of errors. The situation changes after 11:44:20 UT when the temperature
calculated from the model is higher than that calculated from observations and is outside the range of errors. About 11:45 UT there is
a short episode when the temperatures determined from the model and observations are almost
identical.
\begin{figure*}[ht!]
\includegraphics[angle=90,scale=0.6]{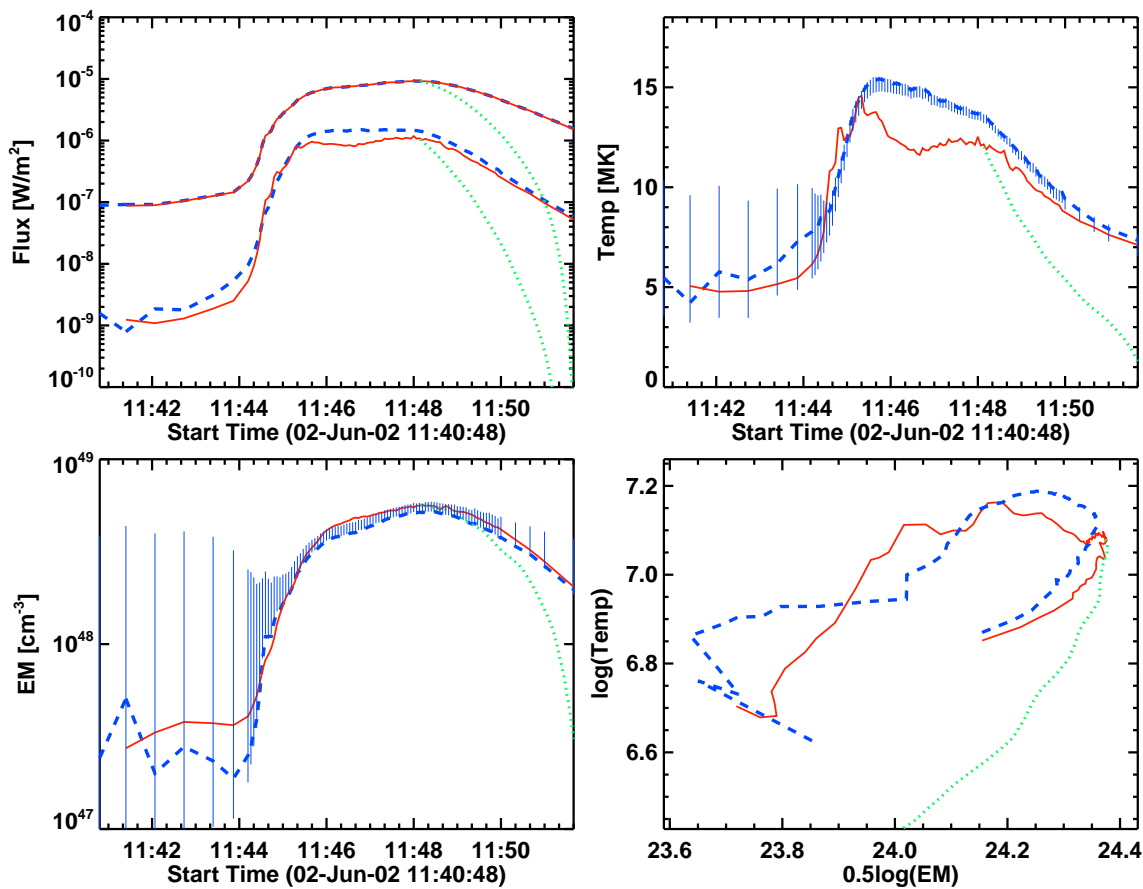}
\vspace{-0.5 cm}
\caption{Results of the modelling of the June $\rm 2^{nd}$, 2002 solar flare. The upper
left panel presents the observed and calculated \emph{GOES} fluxes in 0.5-4 \AA\ (lower
curve) and 1-8 \AA\ (upper curve) energy bands. The upper right panel shows
temperature, lower left panel emission measure, and the lower right panel
presents the diagnostic diagram - log(T) vs. 0.5log(EM). Colour code as used in the
described panels in Figure: blue dashed lines relate to the value observed or calculated
from \emph{GOES} fluxes, red lines show the value calculated from the solar flare
model, green (dotted) lines describe the value calculated from the solar flare
model without any heating after 11:48:08 UT.}
\end{figure*}
After 11:45:20 UT  the highest difference between the temperatures calculated from observations and
from the model is observed. At about 11:46:20 UT the difference is at its greatest and is about 3
MK. After this time the difference gradually decreases to reach a similar value at 11:51:20 UT (again within the range of errors). The
behaviour of the emission measure over time can be seen in Figure 8 (lower left panel). Generally,
it can be said that apart from the beginning, an emissions measure calculated from synthesized
light curves reproduces the course of emission measure calculated from observations very well, and generally their ranges of errors overlap apart from a few discrepancies. A
very useful tool to observe the thermal evolution of the flaring plasma is a diagnostic diagram.
This type of flare thermal analysis was presented for the first time by \citet{Jak92}. On the
diagnostic diagram the emission measure ($\rm 0.5\,log\,EM$) is given on the horizontal axis while
plasma temperature ($\rm log\,T$) is given on the vertical axis. Generally, the evolutionary paths
in the diagrams are very sensitive to the efficiency of the plasma heating during the flare. This
is particularly important and relevant to the cooling phase of the flare when the energy flux is
reflected by the slope of its evolutionary path. During this phase evolution of the flare can occur
in two characteristic limiting branches, firstly, the branch when heating of the flaring loop is
abruptly switched off and in this case, the slope of the evolutionary branch is about 2. This is an
instance of free or nearly-free evolution of the flaring loop, when heating decreases so quickly
that it does not affect the rate of flare cooling.
\begin{figure}[ht]
\begin{center}
\includegraphics[angle=0,scale=0.50]{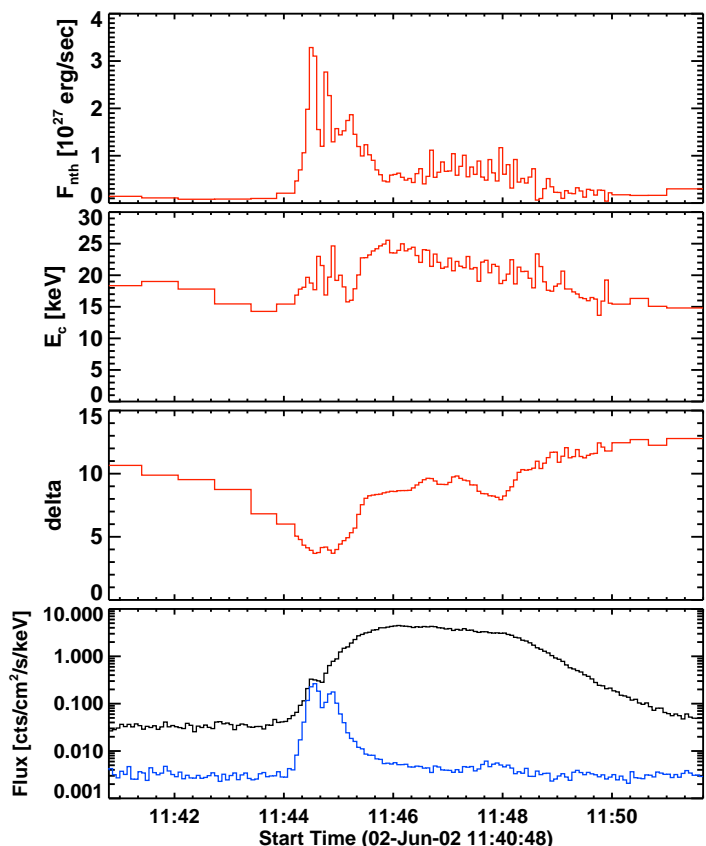}
\end{center}
\vspace{-1.0 cm}
\caption{Time evolution of the thick-target model parameters calculated for
the June $\rm 2^{nd}$, 2002 solar flare using \emph{RHESSI} registered fluxes. From top to
bottom: energy flux of non-thermal electrons $\rm F_{nth}$, low cut-off energy $\rm E_c$, $\delta$
index of energy spectrum, and HXR fluxes: 12-25 keV (black line) and 25-50 keV
(blue line).
}
\end{figure}
Secondly, the quasi-stationary state (QSS) branch when the rate of flare cooling is still fully
covered by the heating and the slope of the evolutionary path in this case is about 0.5. Figure 8
(lower right panel) shows the diagnostic diagram for the June $\rm 2^{nd}$, 2002 solar flare.  The
difference between the observed and modelled evolutionary paths is caused mainly by underestimation
of the modelled temperatures of plasma. During the cooling phase of the flare of June $\rm2^{nd}$,
2002 from 11:48:20 UT to 11:50:20 UT the observed and modelled evolutionary paths have slopes which
are very close to, but less than 2, after the time 11:50:20 UT the slope was about 0.5. This
behaviour indicates that some heating was present during the decay phase of the analysed event. A
green dotted curve shows the evolutionary path calculated in the case when heating of the flaring
loop was abruptly switched off after 11:48:08 UT, and the path reaches a slope very close to 2
after 11:48:50 UT.

\begin{figure*}[ht!]
\includegraphics[angle=90,scale=0.6]{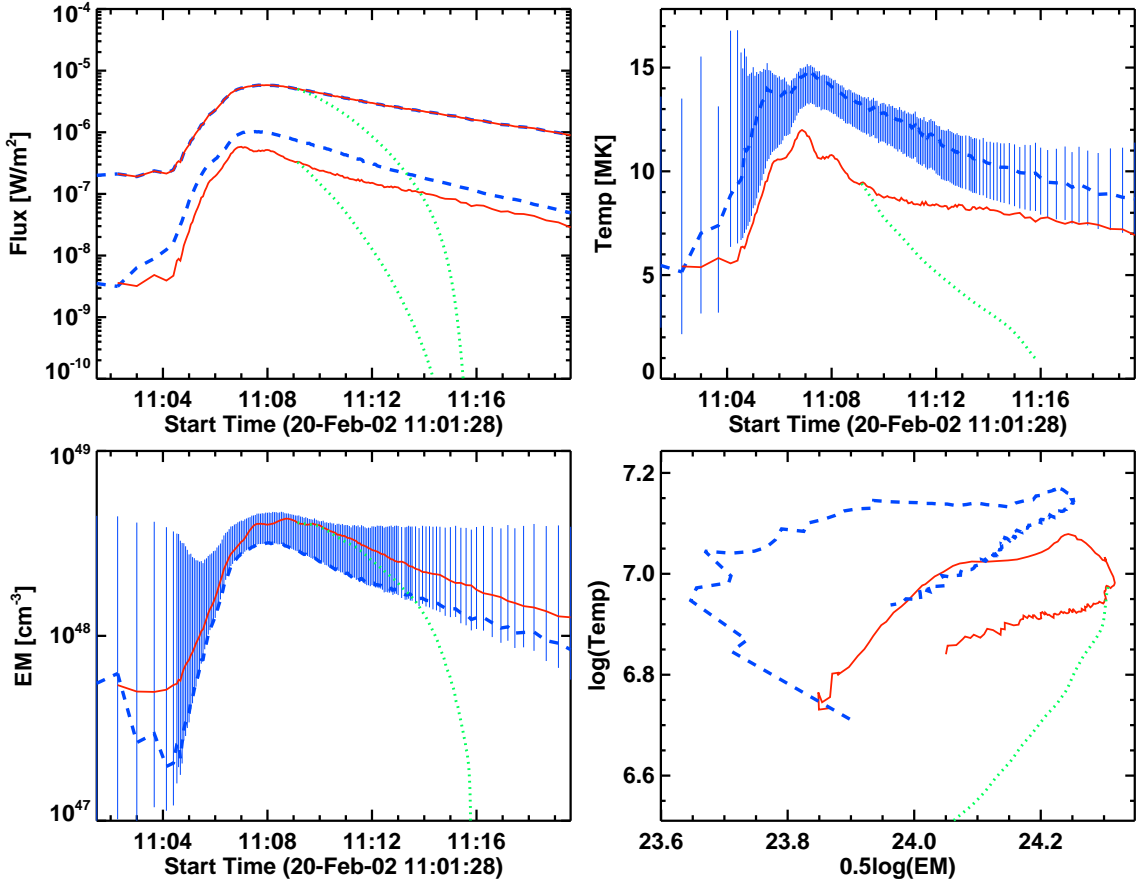}
\vspace{-0.5 cm}
\caption{Results of the modelling of the February $\rm 20^{th}$, 2002 solar flare. The
upper left panel presents the observed and calculated \emph{GOES} fluxes in 0.5-4 \AA\
(lower curve) and 1-8 \AA\ (upper curve) energy bands. The upper right panel shows
temperature, lower left panel emission measure, and lower right panel presents
diagnostic diagram - log(T) vs. 0.5log(EM). Colour code as used in the described
Figure: blue dashed lines relate to the value observed or calculated from \emph{GOES}
fluxes, red lines show the value calculated from the solar flare model, green
(dotted) lines describe the value calculated from the solar flare model without
any heating after 11:08:36 UT.
}
\end{figure*}

Figure 9 presents time variations of the electron beam parameters determined assuming thick target
model and \emph{RHESSI} fluxes in two energy bands, 12-25 keV and 25-50 keV, respectively. As
assuming a already mentioned earlier, non-thermal electron beams characterized by these parameters
provide to the loop energy sufficient to obtain compliance of the observed and modelled \emph{GOES}
fluxes in a 1-8 \AA\ energy band before and during the impulsive phase, and also during the decay
phase of the flare.  The time course of energy flux of NTE $\rm F_{nth}$ varied between $\rm
4.0\times10^{25}$ and $\rm 3.28\times10^{27}$ erg $s^{-1}$, reaching the first maximum at 11:44:40
UT, and the second (strongly blurred) at 11:47:42 UT. These two maxima are reflected well in the
course of the electron spectral index $\delta$, as local minima, when spectrum of non-thermal
electrons was hardened. This behaviour is called in the literature as  soft-hard-soft and most of
the solar flares show during the impulsive phase this pattern of variation of the observed spectral
index \citep[see e.g.,][]{Gri04}. These two maxima are also faithfully reproduced in the light
curves recorded by \emph{RHESSI}, especially in the 25-50 keV energy range where the first maximum
(11:44:40 UT) reflects the main impulsive phase of the flare and a second maximum seen as a small
increase in the flux of 25-50 keV responsible for flattening the SXR maximum observed at the same
time for \emph{GOES} emission and \emph{RHESSI} channels 4-12 keV and 12-25 keV at 11:47:42 UT.
Low energy cut-off $\rm E_c$ varied during the flare between 14 keV and 25 keV.  During the
modelled pre-heating phase of the flare $\rm E_c$ value decreases from 18 keV to 14 keV, then after
several sharp changes increases up to 25 keV during the impulsive phase, and decreases back to 15
keV during the gradual phase of the flare. Temporal variations of the calculated cut-off energy
$\rm E_c$ agree well with estimations of $\rm E_c$ range and temporal variations made in a previous
paper \citep{Fal11} and already presented by various authors, e.g., \citet{Hol03,Sui07} and
\citet{War09a,War09b}.

\begin{figure}[ht!]
\begin{center}
\includegraphics[angle=0,scale=0.50]{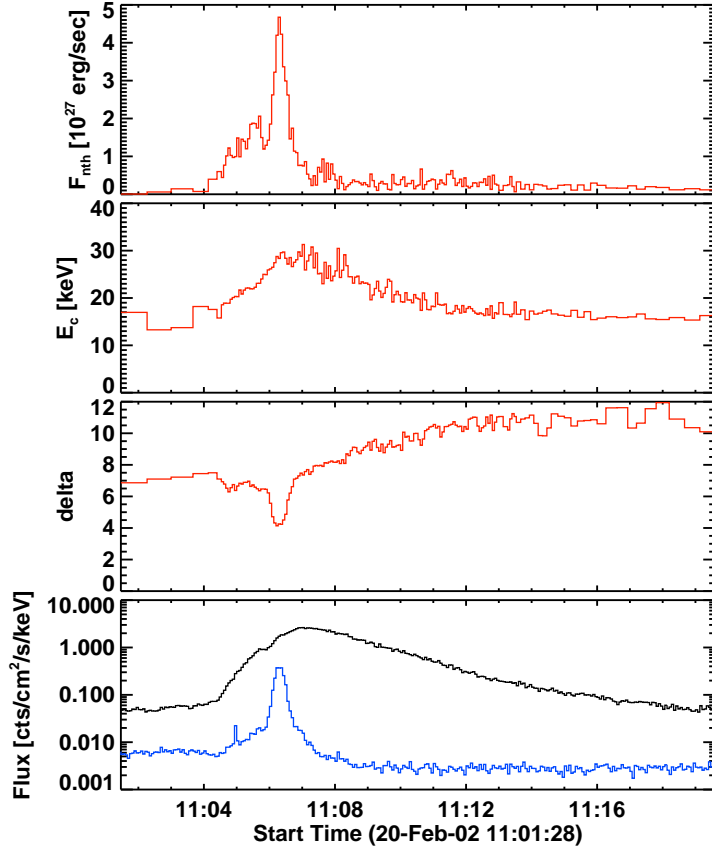}
\end{center}
\vspace{-1.0 cm}
\caption{Time evolution of the thick-target model parameters calculated for
the February $\rm 20^{th}$, 2002 solar flare using \emph{RHESSI} fluxes. From top to
bottom: energy flux of non-thermal electrons $\rm F_{nth}$, low cut-off energy $\rm E_c$,
$\delta$ index of energy spectrum, and HXR fluxes: 12-25 keV (black line) and 25-50 keV
(blue line).
}
\end{figure}

Figure 10 presents the results of modelling of the February $\rm 20^{th}$, 2002 solar flare. In the
case of this event quality of the restoration of \emph{GOES} 0.5 - 4 \AA\ light curve (upper left
panel) is much worse compared to the June $\rm 2^{nd}$, 2002 solar flare. For almost the entire
duration of the analysed flare the modelled light curve (0.5 - 4 \AA) is located below the observed
light curve but an overall pattern of the variations is very similar to the observed ones. The
courses of the light curves show, that during the decay phase some heating occurred. The green
dotted curves in Figure 10 (upper left panel) represent the evolution of the plasma in the flaring
loop without any heating  after 11:08:36 UT and have different courses than the observed ones. A
differences in the courses between synthesized and observed light curves, as in the case of the
first analysed event, also reflects on the calculated temperature and emission measure values. For
temperatures (upper right panel) the biggest difference is observed for the maximum of the flare,
at 11:07:10 UT, when the difference between the temperatures determined from observations and
synthesized light curves reached about 3 MK. After the maximum of the flare the difference between
temperatures gradually decreases to reach a value of about 1.5 MK at 11:19:40 UT. Temperatures determined from
a synthesized light curves only at the beginning and end of the event are within the range of errors of temperatures calculated from the observed light curves.
Time variations
of emission measure are presented in Figure 10 (lower left panel). Emission measure calculated from
the synthesized light curves reproduce the course of emission measure calculated from observation
but observed EMs are higher. For the whole event the course of EM calculated from a synthesized light curve is within the range of errors of EM calculated from the observed light curves.
The main difference between observed and modelled evolutionary paths
on the diagnostic diagram (Figure 10, lower right panel) is caused by underestimation of
temperature and overestimation of emission measure. However, despite the shift both evolutionary
paths are similar. During the cooling phase of the flare from 11:16:20 UT to 11:19:30 UT the
observed and modelled evolutionary paths have a similar slope of about 0.5 and this proves that
some heating was present during the decay phase. The time variation of the parameters of the
electron beam in thick target approximation and \emph{RHESSI} fluxes of the flare are presented in
Figure 11. The energy flux of NTEs $F_{nth}$ varied in time between $5.7\times10^{24}$ and
$4.7\times10^{27}$ erg$ s^{-1}$ reaching a maximum at 11:06:20 UT. Low energy cut-off $\rm E_c$
varied during the analyzing flare between 13 keV and 31 keV and has a similar pattern of variations
to the previously analysed event. Electron spectral index $\delta$ varied between 4 and 12,
exhibiting typical behaviour of variation soft-hard-soft observed during the impulsive phase for
most solar flares. The observed minimum for the electron spectral index is reflected in the maximum
of the light curve recorded by \emph{RHESSI} in the 25-50 keV energy range at 11:06:08 UT.
Previously \citep{Siar09,Fal11}, and in the present work, optimisation of the low energy cut-off of
the electron distribution $\rm E_c$ by comparison of the observed and calculated 1-8 \AA\
\emph{GOES} fluxes is used during the modelling of flares. However, the conformity of the observed
and synthesized light curves in the 0.5-4 \AA\ band serves as a measure of quality of the
calculated flare models. In this approach, these conditions are treated  as independent in energy,
but it is worth noting that the two energy ranges for \emph{GOES} overlap. Therefore, it was
decided to check in this work the accuracy of the model for other energy ranges. The method is
based on a comparison of the synthesized and observed total fluxes in 6-12 keV, 12-25 keV, 25-50
keV and 50-100 keV energy bands. Examples of such comparisons are shown in Figures 12 and 13
respectively for the July $\rm 2^{nd}$, 2002 and February $\rm 20^{th}$, 2002 solar flares. For the
solar flare of June $\rm 2^{nd}$, 2002 (Figure 12) the conformity of the observed and synthesized
light curves in higher energy bands (25-50 keV and 50-100 keV) is very good. In these bands of
energy the largest contribution to the total flux comes from a non-thermal emission. In lower
energy ranges, the contribution of this component decreases and the thermal component begins to
play a dominant role. Observed X-ray light curves are of course the sum of the thermal and
non-thermal emissions. Comparison of the observed light curves with calculated thermal and
non-thermal fluxes, can shows what point in time what type of emission dominated during the various
phases of the flare. In Figure 12 the non-thermal synthesised emission in the energy range 50-100
keV is represented by the dark blue (dotted) curve. For the lower range (25-50 keV) synthesized
emission is represented by the red curves, which are sums of synthesized thermal and non-thermal
emissions, but the contribution of thermal emissions in this energy range is still very small. It
is only in the energy range of 12-25 keV that thermal and non-thermal emissions have comparable
contributions. For example, for this range of energy temporary dominance of one of the components
can be traced. After 11:44:00 UT the non-thermal component dominates, which is associated with the
main impulsive phase of the flare, and at 11:45:10 UT thermal emission begins to dominate up to
11:50:00 UT, where again the two components have similar contributions. The quality of the fit
between an observed and synthesized light curve for the 12-25 keV energy band is worse than at
higher energy ranges. In particular, the differences are apparent after 11:44:20 UT, where the
maximum of the synthesized light curve is lower compared to the maximum of the observed light
curve. The shape of the maximum of the synthesized light curve is restored correctly and can be
seen in the double peaks as in observed light curves.
\begin{figure}[ht!]
\begin{center}
\includegraphics[angle=0,scale=0.58]{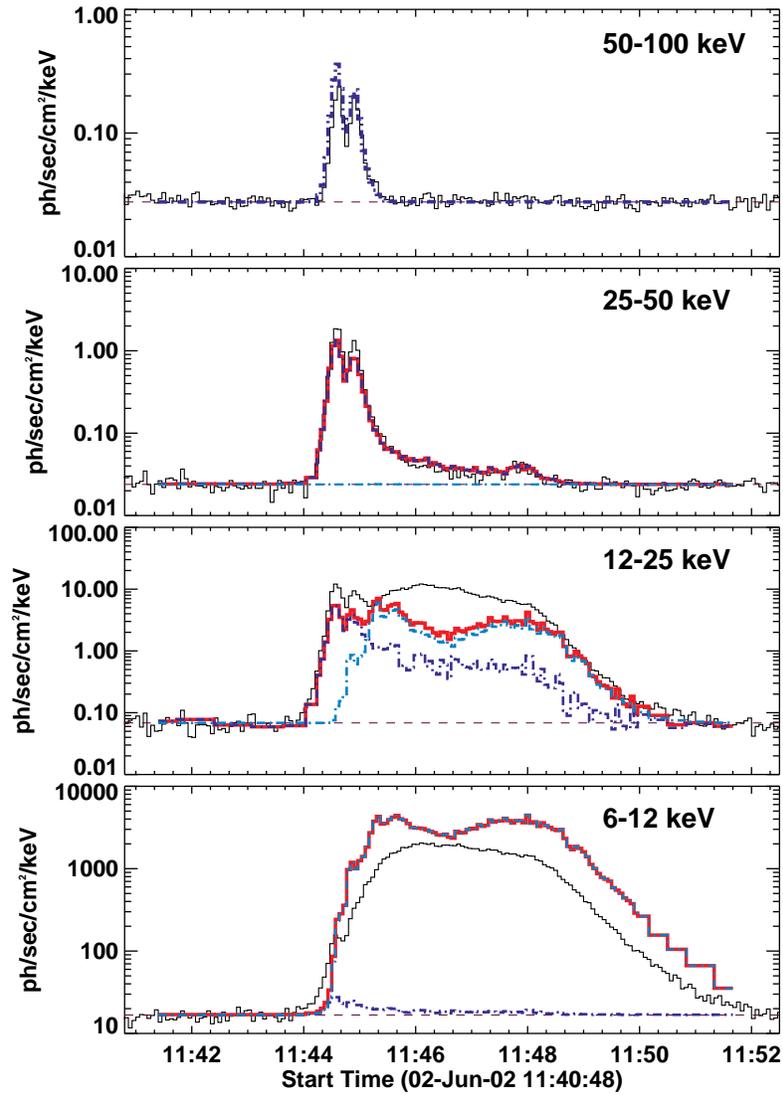}
\end{center}
\vspace{-1.0 cm}
\caption{Results of the modelling of the June $\rm 2^{nd}$, 2002 solar flare. The
figure shows synthesized and observed total fluxes in 50-100 keV, 20-25 keV, 12-25 keV and 6-12 keV energy bands. Colour code as
used in the described Figure: the black curves - observation light curves from
\emph{RHESSI}. The dark blue (dotted) curves - synthesized non-thermal emission. The
bright blue (dotted) curves - synthesized thermal emission. The red curves - sum
of synthesized thermal and non-thermal emissions. The dashed brown line is a
background derived from observations.
}
\end{figure}
\begin{figure}[ht!]
\begin{center}
\includegraphics[angle=0,scale=0.58]{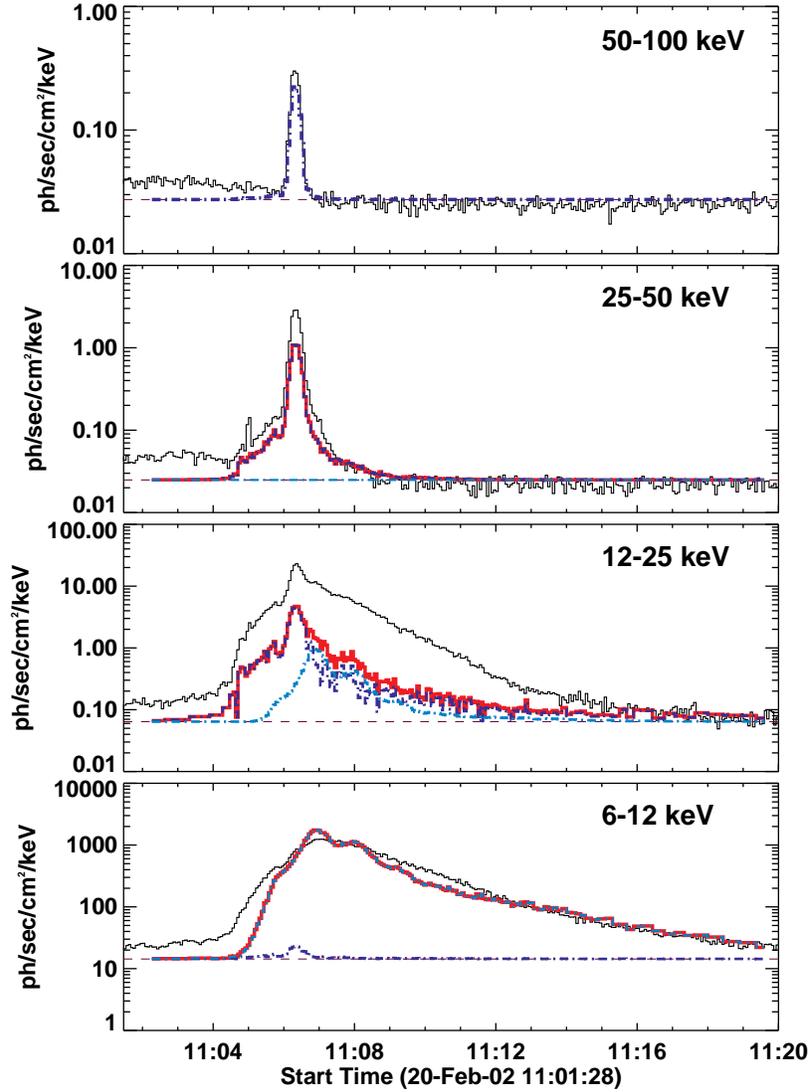}
\end{center}
\vspace{-1.0 cm}
\caption{Results of the modelling of the February $\rm 20^{th}$, 2002 solar flare. The
figure shows the synthesized and observed total fluxes in 50-100 keV, 20-25 keV, 12-25 keV and 6-12 keV energy bands. The black curves - observation light curves from
\emph{RHESSI}. The dark blue (dotted) curves - synthesized non-thermal emission. The
bright blue (dotted) curves - synthesized thermal emission. The red curves - sum
of synthesized thermal and non-thermal emissions. The dashed brown line is a
background derived from observations.
}
\end{figure}

\begin{figure}[h]
\begin{center}
\includegraphics[angle=0,scale=0.65]{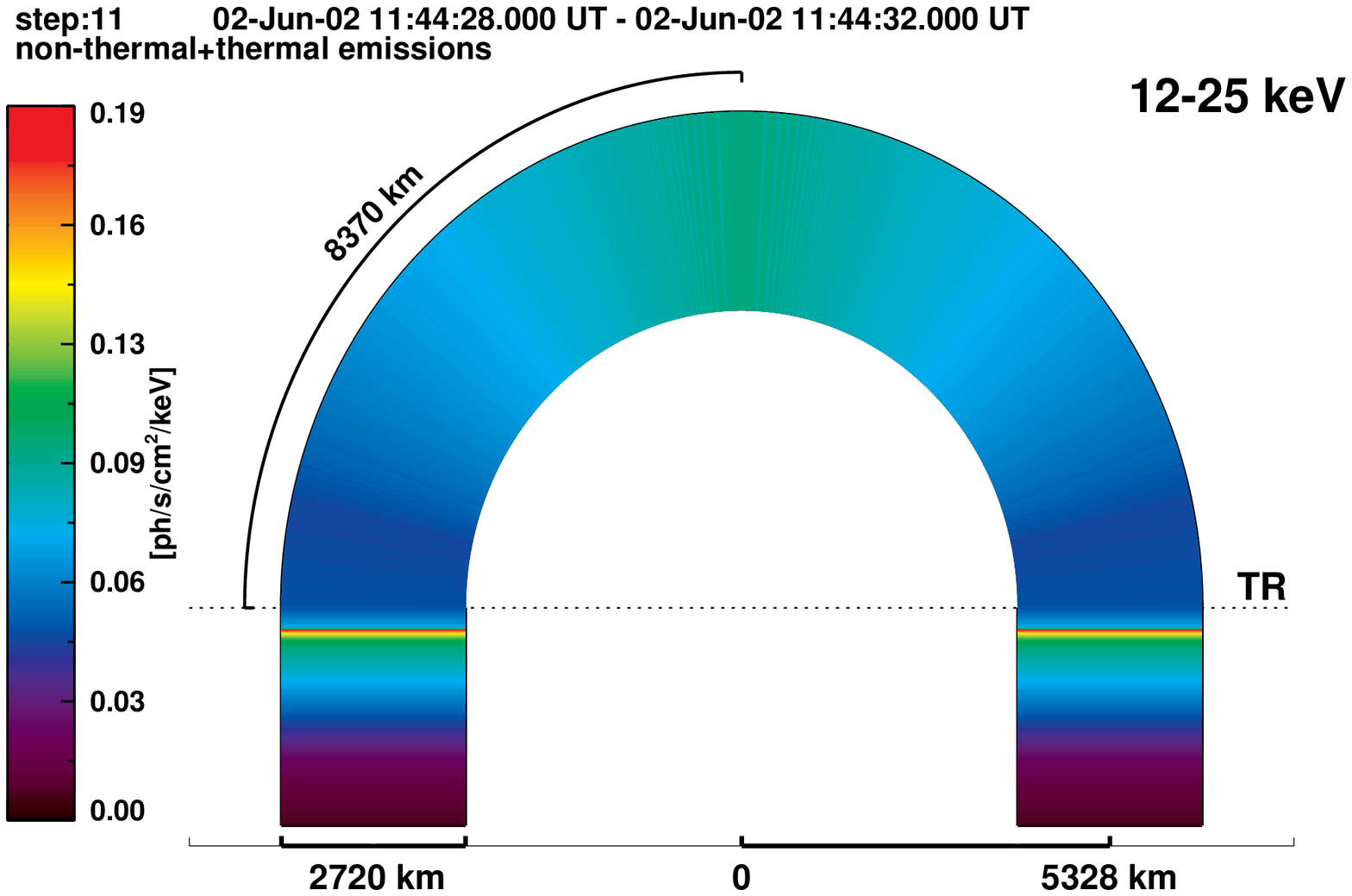}
\end{center}
\vspace{-1.0 cm}
\caption{An example of the simulation of the spatial distribution of integral non-thermal and thermal emission in the energy range 12-25 keV calculated
for a selected evolution stage of the June $\rm 2^{nd}$, 2002 flare. The dashed line represents a level
of the transition region (TR). Impulsive brightenings in the feet of the loop are well
seen while the thermal emission at this time does not contribute (more details in the main text).
}
\end{figure}

After 11:45:10 UT a local minimum appears in the synthesized light curve which does not occur for
the observed light curve. A similar effect at the same time, appeared for the observed light curve
of \emph{GOES} 0.5-4 \AA\ band (see Figure 8, upper left panel). The non-thermal component does not
indicate the described behaviour, therefore this effect corresponds to the thermal component.
Again, after 11:48:30 UT the curves are in a similar course to the end of the simulation. In the
lowest energy channel (6-12 keV) the thermal component dominates, although  during the impulsive
phase the non-thermal component gives a small contribution. The synthesized emission has a similar
pattern of the variation, but runs over the observed light curve. Also in this energy band a local
minimum of the synthesised light curve appears at 11:46:35 UT which does not occur in observed
light curve. The emission calculated from the model in the 6-12 keV begins to clearly start and
grows later and faster than observed one, but this behaviour is not observed in the other described
energy bands.
\begin{figure}[h]
\begin{center}
\includegraphics[angle=0,scale=0.65]{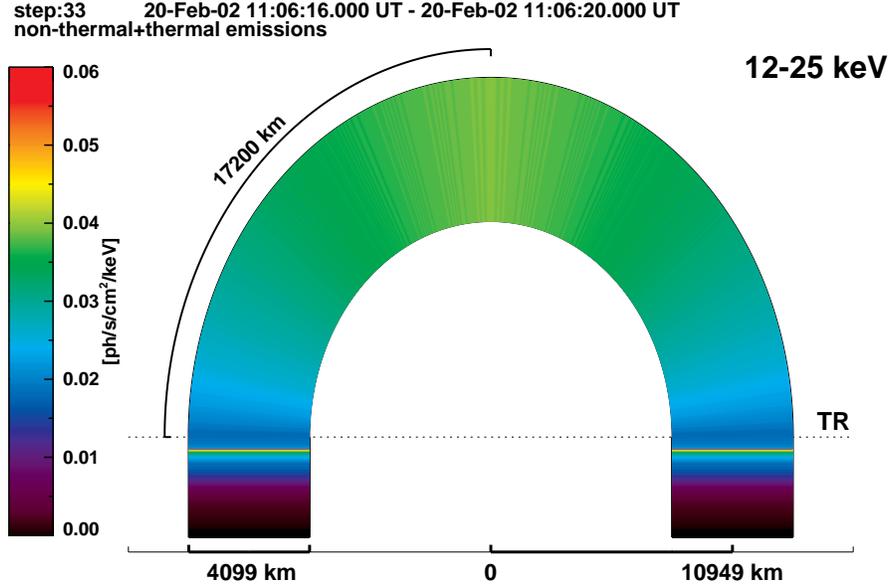}
\end{center}
\vspace{-1.0 cm}
\caption{Simulation of the spatial distribution of total non-thermal
and thermal emission in the energy range 12-25 keV, calculated for a
selected evolution stage of the February $\rm 20^{th}$, 2002 flare. The dashed line represents the level of transition region (TR). Impulsive brightenings in the feet and a bright top of the loop are well seen (more details in the main text).
}
\end{figure}

Figure 13 shows the comparison of the synthesized and observed total fluxes in the same four energy
bands of the February $\rm 20^{th}$, 2002 solar flare. For this event quality of the restoration of
synthesized curves is worse in relation to the June $\rm 2^{nd}$, 2002 solar flare similarly as for
\emph{GOES} 0.5-4 \AA\  light curve. For almost the entire duration of the analysed flare the
modelled light curves are located below the observed light curves, but the variation patterns are
very similar to the observed ones. Only in the 6-12 keV energy band the synthesized curve has a
similar course and values to the observed light curve, but at the beginning, the start of emission
calculated from the model begins clearly later than in observation, just as in the case of the
first analysed flare. The biggest differences between synthesized and observed light curves are in
the energy range of 12-25 keV and they decrease with increasing energy.

Figure 14 presents an example of the simulation of the spatial distribution of integral non-thermal
and thermal emission in the energy range 12-25 keV, calculated for the early-stage of the impulsive
phase of the June $\rm 2^{nd}$,2002 flare. The emission comes mainly from the feet, where impulsive
brightenings appeared caused by non-thermal electrons which heat chromospheric matter. A much
weaker and more diffuse emission comes from the top of the loop, which also has a non-thermal
origin. A more complex picture of the flare is shown in Figure 15. This is an example of the
simulation of spatial distribution of total non-thermal and thermal component calculated also for
the early-stage of the impulsive phase of the February $\rm 20^{th}$, 2002 flare. In this case, as
can be seen in the feet there is strong impulsive brightening and also much weaker and blurry
emission, which is mainly thermal in origin, from the top of the loop.

\begin{table}
\caption{Comparison of observed and modelled fluxes for selected structures of the analysed flares.} 
\label{table:2} 
\centering 
\begin{tabular}{l c c c c c} 
\hline\hline 
\textsc{Date} & & \multicolumn{2}{c}{February $\rm 20^{th}$, 2002} & \multicolumn{2}{c}{June $\rm 2^{nd}$, 2002} \\ 
\hline 
\textsc{accumulation time}         & &\multicolumn{2}{c}{11:06:16 UT (8s)} & \multicolumn{2}{c}{11:44:20 UT (48s)}\\
\hline
& \textsc{units} & \textsc{observations} & \textsc{model}& \textsc{observations} & \textsc{model}\\
\hline
\textsc{energy range}  &  &\multicolumn{2}{c}{35--45 keV} & \multicolumn{2}{c}{30--80 keV}\\
\textsc{isophote 30 \% } &  & & & & \\
\cline{1-1}
$FP_1$         & $ph/s/cm^2/keV$     & 0.68 -- 3.61             & --    & 2.37 -- 6.61   & --      \\
$FP_2$         & $ph/s/cm^2/keV$     & 2.23 -- 6.38             & --    & 1.09 -- 4.41   & --      \\
$FP_{AVE}$     & $ph/s/cm^2/keV$     & 1.46 -- 4.99             & 0.81  & 1.73 -- 5.51   & 7.64    \\
\textsc{box} &  & & &  & \\
\cline{1-1}
$FP_1$    & $ph/s/cm^2/keV$     & 1.72 -- 5.52   & --    &  3.93 -- 9.02& --   \\
 & $arcsecs$ & $\,[^{\,894."49}_{\,911."49}$ $^{,\,263".85}_{,\,282".85}]$& --    & $\,[^{\,-\,149".40}_{\,-\,137".40}$ $^{,\,-\,308".32}_{,\,-\,291".32}]$ & --   \\
$FP_2$    & $ph/s/cm^2/keV$     & 5.29 -- 11.00  & --    &  1.95 -- 5.92& --   \\
 & $arcsecs$& $\,[^{\,891".49}_{\,923".49}$ $^{,\,262".85}_{,\,235".85}]$& --    & $\,[^{\,-\,162".40}_{\,-\,151".40}$ $^{,\,-\,313".32}_{,\,-\,289".32}]$ & --   \\
$FP_{AVE}$& $ph/s/cm^2/keV$     & 3.50 -- 8.26   & 0.81  & 2.94 -- 7.47 & 7.64 \\
$F_{BKG}$ & $ph/s/cm^2/keV$     & 0.06           & --    & 0.07         & --   \\
 & $arcsecs$& $\,[^{\,876".49}_{\,884".49}$ $^{,\,227".85}_{,\,236".85}]$& --    & $\,[^{\,-\,175".40}_{\,-\,165".40}$ $^{,\,-\,334".32}_{,\,-\,325".32}]$ & --   \\
\hline 
\textsc{energy range}  &  &\multicolumn{2}{c}{6--12 keV} & \multicolumn{2}{c}{6--12 keV}\\
\textsc{isophote 30 \% } & \textsc{} & \textsc{} & \textsc{}& \textsc{} & \textsc{}\\
\cline{1-1}
$FL_{NTH}$       & $ph/s/cm^2/keV$     & --               & 66.9   & --               & 289.5    \\
$FL_{TH}$        & $ph/s/cm^2/keV$     & --               & 2710.0 & --               & 11101.1  \\
$FL_{SUM}$       & $ph/s/cm^2/keV$     & 3043.8 -- 3155.2 & 2788.3 & 551.1 -- 503.2   & 11402.9  \\
\textsc{box} &  &  & &  & \\
\cline{1-1}
$FL_{SUM}$       & $ph/s/cm^2/keV$     & 4577.9 -- 4714.2 & 2788.3 & 866.2 -- 926.0   & 11402.9  \\
&$arcsecs$ & $\,[^{\,897".49}_{\,914".49}$ $^{,\,239".85}_{,\,279".85}]$& --    & $\,[^{\,-\,164".40}_{\,-\,137".40}$ $^{,\,-\,315".32}_{,\,-\,292".32}]$ & --   \\
$F_{BKG}$       & $ph/s/cm^2/keV$     & 11.4              & --     & 12.3             & --       \\
 & $arcsecs$& $\,[^{\,875".49}_{\,884".49}$ $^{,\,226".85}_{,\,236".85}]$& --    & $\,[^{\,-\,175".40}_{\,-\,166".40}$ $^{,\,-\,334".32}_{,\,-\,325".32}]$& --   \\
 \hline\hline
\multicolumn{6}{l}{{\footnotesize $FP_1$ and $FP_2$ - flux measured at footpoints; $FP_{AVE}$ -  average flux from $FP_1$ and $FP_2$; $F_{BKG}$ -  background flux;}} \\
\multicolumn{6}{l}{{\footnotesize $FL_{NTH}$ and $FL_{TH}$ - flux of non-thermal and thermal emissions from entire flare's loop; $FL_{SUM}$ - the total flux coming}} \\
\multicolumn{6}{l}{{\footnotesize from the entire flare's loop.}} \\
\end{tabular}
\end{table}

Finally, spatial distributions of the signal in the \emph{RHESSI} images were compared with
emissions resulting from the model. Using the spatial distribution of the observed signals
presented in Figures 2 and 5 (left panels) fluxes were measured for the selected structures and
compared with the values obtained from the models for both analysed flares. The results are
presented in Table 2. For selected energy ranges 35-45 keV and 30-80 keV emissions have a
non-thermal origin, which can be proven by looking at the total fluxes in Figures 12 and 13 for
June $\rm 2^{nd}$, 2002 and February $\rm 20^{th}$, 2002, respectively. On the \emph{RHESSI} images
emissions originate chiefly from the feet of the flaring loops. For this reason, in the images were
selected areas of the feet using the two connected methods. The first one delimits the interest
area using the box defined as coordinates of the corners (see Table 2), the second delimits the
feet with an isophote of 30\% intensity of the brightest pixel in this box. In this way the areas of
footpoints were estimated in the images of both analysed events. However, the observed fluxes
of the footpoints are not identical (see Table 2 - $FP_1$ and $FP_2$). Therefore, taking into
account the symmetry of the numerical model, the corresponding parameter was calculated as an
average of the fluxes of both feet ($FP_{ave}$). The non-thermal footpoint flux is calculated for
the model  using formula~(\ref{eq1}) and additionally from
$F_{FP}=\int_{\varepsilon_1}^{\varepsilon_2}\int_{s_{TR}}^{s_{MAX}}
I(\varepsilon,s)\,ds\,d\varepsilon$ as total intensity below transition region ($\rm s_{TR}$) and
integrated in selected energy ranges ($\varepsilon$). For these calculated fluxes added value of
the background was derived from the \emph{RHESSI} image ($F_{BKG}$) using the box defined as
coordinates of the corners (see Table 2). For the lowest energy range (6-12 keV) the
applied methodology is more complicated. In this energy range and at the time when the images were
registered for the analysed flares the thermal component dominates, although the non-thermal
emission adds a small contribution (see Figures 12 and 13). Therefore, calculated emission of the
model in this energy range should be the sum of the thermal ($FL_{TH}$) and non-thermal
($FL_{NTH}$) emissions and such value ($FL_{SUM}$) is compared with the observations. The modelled
emission was integrated over the entire loop, and also in this case the value of the background is
added and derived from the \emph{RHESSI} image. The only observed structure of the \emph{RHESSI}
images in this energy range is the flare's loop. As in the case of a high energy range the same two
methods of estimation of the areas of loop's emission were used.

For the analysed event of June $\rm
2^{nd}$, 2002 in the 30-80 keV energy range emission from the model is about 2 times bigger than
observed flux within an isophote of 30\% and is outside the range of errors. A similar situation occurs
for the observed flux using a box method, where the obtained the synthesized emission is 30\% higher than the emission derived from the observation. A much larger inconsistency occurs for the 6-12 keV energy range. In this
case the emission calculated from the model is about 20 times greater within an isophote of 30\% and
about 12 times greater using the box than the observed emissions, respectively. The obtained result
generally agrees with the relations of the synthesized and observed total fluxes shown in Figure 12
for the selected time. Results of photometry of the June $\rm 2^{nd}$, 2002 solar flare show that
the synthesized emission in the 35-45 keV energy range is about 4 times lower than the observed
emission within an isophote of 30\%. Within the defined box this ratio is about 7. Good agreement was
obtained for the 6-12 keV energy range, where emission from the model is about 11\% lower for
an isophote of 30\%, and about 60\% lower for the box than the observed emission. As for the first
analysed event, the received result agrees with the relations of the synthesized and observed total
fluxes shown in Figure 13 for the selected time.

\section{Discussion \& Conclusions}

The modelling of the two solar flares observed on February $\rm 20^{th}$, 2002 and June $\rm
2^{nd}$, 2002 was conducted under the assumption that only non-thermal electron beams delivered
energy to the flaring loop via the Coulomb collisions and the NTEs heating function was calculated
based on the parameters derived from observed HXR spectra. Comparison of modelling results with
observations shows that SXR and HXR emissions of both single-loop like events could be fully
explained by electron beam-driven evaporation only. There is no need for any additional heating
mechanism to explain the observed SXR and HXR emissions and dynamics of the flaring plasma. The use
of the Fokker-Planck formalism allowed for an improvement of the calculated models, if compared
with previous works, by direct comparison of the observed and modelled distributions of the HXR
emission. The best compatibility of the model and the observations was obtained for the June $\rm
2^{nd}$, 2002 event, both the 0.5-4 \AA\ \emph{GOES} light curve and for total fluxes in 50-100
keV, 25-50 keV, 12-25 keV and 6-12 keV energy band. Poorer concordance compared to the first event
is obtained for the February $\rm 20^{th}$, 2002 solar flare. In both cases, the largest
differences between the observed and synthesized light curves appeared in the 12-25 keV energy
range.
This difference is mainly caused by a thermal emission component which gives too small a
contribution in this range to the total flux. In higher ranges of energy, differences between the
model and the observed fluxes decrease. In these ranges the non-thermal component dominates. This
behaviour appears to be justified because, even if the plasma distribution in the model does not
correspond to the observation, the non-thermal electrons which have high energies can penetrate
deeper into the chromosphere and do not interact with matter during their travel in contrast to the
low energy non-thermal electrons that convey energy in the higher parts of the flare loop. The HXR
non-thermal emission is directly related to the flux of the accelerated NTEs whereas the SXR and
HXR thermal emissions are related to the energy deposited by the same NTEs flux. Another
important factor influencing the differences between the modelled and observed fluxes appearing in the
12-25 keV energy range is the assumption that injected electrons have a power law energy spectrum.
As shown by \citet{Liu09} using the Fokker-Planck model of particle transport and the
injected electron spectrum based on stochastic acceleration, higher coronal temperatures and
densities, larger upflow velocities, and faster increases of these quantities are obtained than in
the model where electron injection spectrum was of power law. This is due to the fact that the
injected electron spectrum smoothly spans from a quasi-thermal component to a non-thermal tail. The
injected beam of electrons contains a lot of low energy electrons which deposit their energy
primarily in the corona and indirectly cause stronger evaporation, and thereby the thermal emission
component could give a greater contribution to the total flux in the 12-25 keV energy range. An
underestimation of the thermal component in the model may also probably indicates inconsistencies
between distributions of the mass in real flaring
loops and in their models. This wrong distribution of mass in the model may arise by relative
simplicity of the applied numerical 1D code, coarse estimation of the geometrical loop parameters,
errors in \emph{RHESSI} spectra restoration and beam parameters obtained from fitting NTEs,
possible problems in a \emph{GOES}'s calibration and by using an analytical formula for NTEs
heating function. Additionally, it is assumed that at the beginning we have an initial,
quasi-stationary preflare model of the flaring loop with the mass distribution consistent with the
observations and begins to heat the plasma using NTEs heating function which was calculated based
on the parameters derived from observed HXR spectra. These parameters are determined by fitting
\emph{RHESSI} spectra and vitiated by errors.
The parameters uncertainties reported by OSPEX can give some idea of the stability of the
fit results. They do not account for interdependence between fit parameters (e.g. between the
temperature and emission measure values of an isothermal emission component or the power-law index
of the electron energy distribution, and low energy cut-off of the electron distribution for the
non-thermal component) and can often underestimate the true uncertainties when the chi-squared
space is not well-behaved (e.g. when it is degenerate along one or more dimensions). Additionally,
an influence of the thermal component on the spectra can be a source of errors of the
derived parameter of electrons' distribution. A detailed analysis of this type of relations and their
influences on the derived parameters of the fitted spectra was conducted by \citet{Saint05}. In
addition, the use of analytical formula for NTEs
heating function, which is only an approximation, may introduce additional errors into the model.
However, taking into account the results published by \citet{Liu09} that the
model with analytical approximation of heating gives about 10\% difference only, compared to the
model with a Fokker-Planck regime, this is an acceptable approximation especially when we
consider the benefits of the speed of calculation and all the other errors arising from the
simplifications used in the model. As a result of all these factors the energy will be deposited
in different layers of the flare model compared to the real event. This may lead to receiving other
physical parameters of the plasma evolution in the loop model than calculated from observations,
particularly the distribution of mass in the flaring loop. This causes the previously described
differences between the observed and synthesized light curves. One can also assume that at the
beginning, an initial, quasi-stationary preflare model of the flaring loop with mass distribution
inconsistent with the observations which may be caused by poorly determined geometrical parameters.
In this case, even the use of the parameters derived from observed HXR spectra without errors and
``ideal'' NTEs heating function will lead to the same effects as in the first case. In reality,
however, the analytical heating function is used  (with all its consequences) and its parameters
with errors.
Similarly the results of photometry for a high energy range show that the best
compliance was obtained for the June $\rm2^{nd}$, 2002 flare where the synthesized emission in the
best case is only 30\% higher than the emissions measured from the observation. For the February $\rm 20^{th}$,
2002 flare, synthesized emission is about 4 times lower than the observed one. In the low energy range
the best conformity was obtained for the February $\rm 20^{th}$, 2002 flare, where emission from the model
is only about 11\% lower than the observed one. The larger inconsistency occurs for the June $\rm2^{nd}$,
2002 solar flare, where synthesized emission in the best case is about 12 times more than
emission from observation. The observed differences which are evident on conducted photometry are
primarily caused by the assumption that the model of the flare is symmetric and there are no differences
in the emissions originating from the feet of the flare's loop. In fact, as has been demonstrated already, the
observations indicate that this condition is not preserved (see Table 2).  The obtained results are also
affected by all the factors that were presented in an earlier discussion.

For instance the obtained errors are similar to results from previous work
\citep{Fal11} and for the non-thermal fits in the decay phase had small formal errors of fitting
parameters (about 1\% for $\delta$ and $\rm E_c$), while the statistical errors were of the order
of 15-25\% for $\delta$ and 11-19 \% for $\rm E_c$. The error ranges were determined for each of
the analysed events. The problem of errors may be particularly important for the very early and
decay phases of a solar flare where the spectra registered by \emph{RHESSI} could also be
reasonably fitted using a thermal model only with an acceptably low value of $\chi^2$ estimator. In
these cases, the temperature values obtained were higher than the temperature calculated from the
\emph{GOES} light curves at the same time. For the early phase this difference is so high that
almost certainly a non-thermal component in the spectra occurred. This situation is not so clear
for the decay phase. Differences of temperature between the \emph{RHESSI} and \emph{GOES} are not
so large, and amount to about 2-3 MK only. In this phase of the flare the thermal component
dominates and covers the weak non-thermal component in the spectra. The argument for the occurrence
of non-thermal emission at this stage of the flare can be extended only on the basis of the model
where during the decay phase heating is necessary (see the green dotted curve in Figure 8 and 10
(upper left panel) representing the evolution of  the plasma in the flaring loop without heating).
Another argument might be an idea presented by \citet{Bros09} assuming that during this phase
non-thermal emission can be below \emph{RHESSI}'s detection threshold.

Despite all these problems, flare modelling presented in this paper simulates the main physical
processes in the right way. It is worth stressing that the HD-1D codes are sufficient in modelling
of the flaring events occurring in simple, single-loop like configurations of the magnetic field,
even in such elaborated applications like a synthesis of the BCS spectra \citep{Fal09b}. Comparison
of the synthesized and observed spectra confirm that the distribution of matter and the velocities
in the flaring loop agree roughly with the observed values.

In summary, as shown in our previous papers \citep{Siar09,Fal11}, efficient heating of the flaring
plasma by non-thermal electron beams starts in the very early phase of the flare, significantly
before the impulsive phase, and lasts a long time during its gradual phase. This heating is so
efficient that no other heating mechanism is necessary to explain emitted SXR and HXR fluxes. In
particular, during the gradual phase of the flare a much slower decay of the SXR flux is observed
than could be expected from an analytical estimation of energy losses of the flaring plasma by
radiation and by conduction \citep{Car95} or from HD numerical modelling of the flares \citep[for
example:][]{Reale97}. However, the SXR decay time of the solar flares is usually much extended if
compared with the previously mentioned cooling times, proving that during this phase of the flares
significant heating is also present. Similar conclusions have already been shown in papers by \citet{McTier93} and \citet{Jiang06}.
 As demonstrated in the present work for the analysed $\rm
20^{th}$ February, 2002 and $\rm 2^{nd}$ June, 2002 flares all of the energy necessary for heating
the flaring loops during all phases of the events was supplied by the NTEs. However, please note
that this does not preclude the existence of other, secondary sources of heating or mechanisms of
energy transfer.

During modelling the flares, the light curves for a number of SXR and HXR energy ranges were
successfully reproduced.  The use of the extended code with the Fokker-Planck formalism allowed for
an improvement of the calculated models and made it possible to ascertain which components of the
emission (thermal or non-thermal) play a dominant role in the chosen energy range. However, taking into account the results of photometry, no doubt a significant refinement of the applied numerical models and more sophisticated implementation of the various physical mechanisms involved are required to achieve a better agreement.

\section{Acknowledgments}
The author acknowledges the \emph{RHESSI}, and \emph{SOHO} consortia for the excellent data and is
grateful to an anonymous referee for constructive comments and suggestions, which have proved to
be very helpful in improving the manuscript. The research leading to these results has received
funding from the European Community's Seventh Framework Programme ([FP7/2007-2013]) under grant
agreement n$^\circ$ [606862].


\end{document}